\IfFileExists{elsarticle.cls}{%
  \documentclass[3p,times,number,nopreprintline]{elsarticle}
  \newif\ifelsevierclass
  \elsevierclasstrue
}{%
  \documentclass[11pt]{article}
  \newif\ifelsevierclass
  \elsevierclassfalse
}

\def\journallayout{}

\usepackage[T1]{fontenc}
\usepackage[utf8]{inputenc}
\usepackage{amsmath,amssymb}
\usepackage{graphicx}
\usepackage{booktabs}
\usepackage{array}
\usepackage{tabularx}
\usepackage{xltabular}
\usepackage{placeins}
\usepackage{algorithm}
\ifdefined\journallayout\else\IfFileExists{geometry.sty}{\usepackage[margin=1in]{geometry}}{}\fi
\IfFileExists{setspace.sty}{\usepackage{setspace}}{}
\usepackage{xcolor}
\usepackage{hyperref}

\ifelsevierclass
\else
  \usepackage[numbers,sort&compress]{natbib}
  \newenvironment{frontmatter}{}{}
  \newenvironment{keyword}{\par\noindent\textbf{Keywords: }}{\par}
  \newcommand{\sep}{; }
  \newcommand{\journal}[1]{}
\fi

\newenvironment{revision}{}{}

\newenvironment{nthreerev}{}{}

\newenvironment{genrev}{}{}
\newcommand{\gencell}[1]{#1}

\definecolor{r4blue}{HTML}{1a56a8}

\definecolor{e1teal}{HTML}{0f6f5c}
\newenvironment{e1rev}{}{}

\definecolor{n16red}{HTML}{c1121f}
\newenvironment{n16rev}{}{}
\newcommand{\nsixteencell}[1]{#1}

\newcommand{\newrev}[1]{#1}

\newcommand{\mrrev}[1]{#1}

\newcommand{\drev}[1]{#1}

\newcommand{\crev}[1]{#1}

\newcommand{\frev}[1]{#1}

\definecolor{n128red}{HTML}{c1121f}

\newcommand{\smsec}[1]{\ref{sm:#1}}

\newcommand{\skelfig}[4][\linewidth]{%
  \begin{figure}[!tbp]
    \centering
    \IfFileExists{#2}{%
      \includegraphics[width=#1,height=0.86\textheight,keepaspectratio]{#2}%
    }{%
      \fbox{\parbox[c][0.22\textheight][c]{0.88\linewidth}{\centering\small Pending rendered figure:\par\nolinkurl{#2}}}%
    }
    \caption{#3}\label{#4}
  \end{figure}
}

\IfFileExists{needspace.sty}{\usepackage{needspace}}{\providecommand{\needspace}[1]{}\providecommand{\Needspace}[1]{}}

\makeatletter
\g@addto@macro\normalsize{%
  \setlength{\abovedisplayskip}{7pt plus 2pt minus 3pt}%
  \setlength{\belowdisplayskip}{7pt plus 2pt minus 3pt}}
\makeatother

\hypersetup{
  hidelinks,
  pdftitle={PINN-Phase: Physics-informed neural integration for accurate, generalizable, and scalable multiphase-field evolution},
  pdfauthor={Seifallah Elfetni; Sascha Thinius; Peter H. Seeberger; Theodore Tyrikos-Ergas},
  pdfsubject={Physics-informed neural integration for accurate, generalizable, and scalable multiphase-field evolution},
  pdfkeywords={physics-informed neural networks; multiphase-field modeling; grain growth; microstructure evolution; virtual microstructure screening; topology change},
  pdfcreator={LaTeX with elsarticle},
}
\ifelsevierclass
  \makeatletter
  \def\ps@pprintTitle{\let\@oddhead\@empty\let\@evenhead\@empty
    \let\@oddfoot\@empty\let\@evenfoot\@empty}
  \makeatother
\fi

\begin{document}
\IfFileExists{setspace.sty}{\singlespacing}{}

\begin{frontmatter}
\title{PINN-Phase: Physics-informed neural integration for accurate, generalizable, and scalable multiphase-field evolution}
\ifelsevierclass
  \author[ctc]{Seifallah Elfetni\corref{cor1}}
  \ead{seifallah.elfetni@ctc-germany.org}
  \author[ctc]{Sascha Thinius}
  \author[ctc,mpikg]{Peter H. Seeberger}
  \author[ctc]{Theodore Tyrikos-Ergas\corref{cor1}}
  \ead{theodore.tyrikos-ergas@ctc-germany.org}
  \cortext[cor1]{Corresponding authors.}
  \address[ctc]{Center for the Transformation of Chemistry, Puschstr.~6b, 04103 Leipzig, Germany}
  \address[mpikg]{Max Planck Institute of Colloids and Interfaces, Biomolecular Systems Department, Am Muehlenberg~1, 14476 Potsdam, Germany}
\else
  \author{Seifallah Elfetni, Sascha Thinius, Peter H. Seeberger, Theodore Tyrikos-Ergas\\Center for the Transformation of Chemistry}
  \date{}
\fi

\begin{abstract}
The value of polycrystal phase-field simulations for materials exploration depends on whether individual grains, not only aggregate statistics, evolve correctly in time. We introduce PINN-Phase, a neural time integrator that advances the multiphase field autonomously and enforces phase bounds and unit sum at every step. It learns from physical targets evaluated on its own predicted states, without reference trajectories in training. A hybrid realization meets all predefined criteria with fixed weights on seven of eight prospective unseen two-dimensional microstructures and both stress cases. It also keeps the exact survivor set in five of six unseen $128^3$ microstructures over 24,000 steps. A separately trained phase-local realization with 2,160 parameters, trained at $48^3$ with eight grains, runs with unchanged weights on microstructures with up to 5,000 grains. With 1,000 and 5,000 grains it reproduces 4,540 of 4,543 reference grain extinctions by identity within their timing tolerance; the other three straddle the evaluation horizon. Typical per-grain volume errors stay nearly constant from 150 to 5,000 grains, while a few large grains later exceed the volume criterion. Time-rescaling controls show that statistics compared at matched coarsening states cannot detect a $3\%$ clock error. In a virtual isothermal hold, that error ends the hold prematurely in all 34 decisions, whereas PINN-Phase stays within $0.3\%$ of the reference time to target. These results hold within the tested microstructure families.

\end{abstract}

\begin{keyword}
physics-informed neural networks \sep multiphase-field modeling \sep grain growth \sep microstructure evolution \sep virtual microstructure screening \sep topology change
\end{keyword}
\end{frontmatter}

\begin{revision}
\section{Introduction}
\label{sec:intro}

Microstructure evolution links processing to the mechanical and functional response of
polycrystalline materials. Grain boundaries migrate under capillary forces, domains coarsen, and
small grains disappear, continually changing microstructure topology. These processes
underlie classical descriptions of normal grain growth and modern full-field studies~\cite{hillert1965,kamachali2015geometrical,kamachali2010multiscale,maire2016,li2022revealing}.
Phase-field (PF) models are a natural continuum framework: diffuse order-parameter
fields represent interfaces without explicit front tracking, and topological events emerge
from the evolution equations~\cite{cahn1958free,allen1979microscopic,steinbach2009,provatas2010,chen2002phase,abrivard2012phase}.
In multiphase-field (MPF) formulations, separate phase fields represent individual grains or
phases, coupled through local phase-fraction constraints~\cite{steinbach1996phase,kamachali2012}.
MPF is well established in two and three dimensions and can incorporate anisotropic
boundary properties, stored-energy gradients, particles, and external fields~\cite{miyoshi2017multi,rezaei2022multi,schwarze2016phase,grose2022multi,moelans2022anisotropic};
level-set formulations offer a complementary full-field route to grain growth~\cite{fausty2018levelset}.
The computational cost of such full-field models, however, grows rapidly with resolution, simulation time, phase
count, and dimension, and large parameter studies repeat it for many related initial
conditions. Adaptive discretizations alleviate this burden but do not change that structure~\cite{tourret2022,li2012amr,xu2022multilevel}.
When such simulations are used to explore processing routes, the quantities of interest include
which grains survive, how large they grow, and when the population reaches a given state.

Learned surrogates offer a complementary route when many related forward evaluations are required:
recurrent latent-space models, conventional machine-learning surrogates, and neural operators have
accelerated phase-field predictions by learning from simulated trajectories~\cite{hu2022accelerating,montes2021accelerating,oommen2022learning}. Recent examples
include resolution-invariant grain-growth operators~\cite{peivaste2025fno},
hybrid adaptive convolutional operators~\cite{bonneville2025hybrid},
operator-network surrogates for solidification grain growth~\cite{ciesielski2025deeponet},
energy- and numerics-consistent operators for 3D phase
fields~\cite{bamdad2026penco}, and graph-recurrent frameworks for long-term 2D and
3D microstructure forecasting~\cite{razavi2026gcnlstm}.
\newrev{Related developments for Allen--Cahn dynamics include fully-discrete operator
learning~\cite{geng2024fullydiscrete} and energy-dissipation-preserving residual
training~\cite{kutuk2025energy}. These approaches differ in supervision budget, state
representation, governing operators, discretizations, training data, and evaluation trajectories,
so no quantitative comparison across them is attempted here.}

Physics-informed neural networks (PINNs) replace or augment trajectory supervision with a residual
of the governing equations~\cite{raissi2019,karniadakis2021,cuomo2022}. Most PINNs use coordinate
networks and collocation losses; long-time integration and constrained gradient flows have motivated
causal training, temporal decomposition, adaptive sampling, time marching, and energy-based
strategies~\cite{daw2023mitigating,penwarden2023causal,meng2020ppinn,jagtap2020xpinn,chen2024atpinn,wu2023sampling,guo2024energy}.
Phase-field applications include Allen--Cahn and Cahn--Hilliard systems, fracture, damage, and
corrosion~\cite{wight2020solving,chen2023pfpinns,goswami2020transfer,zhang2023robust,rojas2023damage,chen2025sharppinns}.
Convolutional-recurrent physics-informed solvers have also advanced neural fields autoregressively
with boundary conditions encoded in the architecture~\cite{REN2022114399}, and physics-informed
networks have served as surrogates for process optimization in materials
manufacturing~\cite{wurth2023pinn}.

\newrev{For multiphase and microstructure modeling, PINNs-MPF uses phase-wise coordinate networks
with space, time and phase decomposition and soft phase-sum enforcement~\cite{elfetni2025pinnsmpf}.}
\mrrev{A preliminary study~\cite{elfetni2025preprint} reported a precursor hybrid architecture in an
energy-based transfer-learning setting.} \newrev{GrainGNN represents the microstructure by graph
features and explicit event operations learned from phase-field trajectories~\cite{qin2024graingnn},
and physics-embedded graph models target process-specific microstructure
prediction~\cite{xue2022physics}. PF-PINNs and Sharp-PINNs address coupled Allen--Cahn/Cahn--Hilliard
corrosion~\cite{chen2023pfpinns,chen2025sharppinns}.} \mrrev{Phase-Field DeepONet and PINO
formulations learn phase-field evolution operators from energy or governing-equation
residuals~\cite{li2023phasefielddeeponet,gangmei2025pino,chen2026pfpino}.} \newrev{PFNet and PENCO
combine trajectory supervision with physics-informed architectural or loss
terms~\cite{xiong2026pfnet,bamdad2026penco}.}

Learned models of three-dimensional grain growth now reach large scales. 3D-PRIMME, the
three-dimensional extension of PRIMME~\cite{yan2022primme}, learns a local update rule from
consecutive frames of a stochastic mode-filter simulation. With unchanged weights, it runs on
domains with up to about 550,000 grains. In direct comparisons with the reference model, its
mean-size kinetics and face statistics agree closely; its statistics remain consistent across domain
sizes, and inclination-dependent growth is learned from the data~\cite{tian2026primme}.
Individual-grain comparisons against direct simulation have been shown for representative grains in
process-specific settings~\cite{xue2022physics}. This progress makes a complementary question
important for numerical microstructure screening: is the evolution of individual grains accurate in
both trajectory and timing? The answer is not automatic. A model whose kinetics run on a slightly
wrong clock passes through the correct sequence of states. Statistics compared at the same
coarsening state then agree, yet every grain has the wrong size at a given time.

A learned time integrator for many-grain multiphase-field evolution must therefore meet several
requirements at once. It must preserve local phase admissibility and remain stable as the active
grain set changes. It must represent coalescence and extinction without accumulating per-step errors
over long rollouts, because high voxelwise agreement can otherwise coexist with a wrong survivor set
or a mistimed extinction. We ask whether one autoregressive framework can meet these requirements. It
should advance the full $N$-component field, learn from a physical residual evaluated on its own
state rather than from reference trajectories, and enforce the phase-fraction constraints after every
update. It should also retain grain identities, so that every extinction and grain volume can be
checked against the reference. The question has four parts, which organize the paper. They are foundation, generalization to
unseen microstructures and longer horizons, and scalability to larger domains and more grains with
unchanged weights. The fourth is reliability: which observables establish fidelity grain by grain and
in time, and what a timing error means for a decision.

PINN-Phase is a physics-informed neural time integrator for constrained multiphase-field evolution.
The hybrid realization evaluates a projected multiphase-field residual on its own predicted state. A
site-local multilayer-perceptron branch represents the local response of the diffuse interface, and
a convolutional-recurrent branch carries spatial interactions and temporal memory. Their combined
response becomes a bounded, phase-sum-neutral increment, and a clip-and-renormalize map restores the
phase bounds and unit sum of the carried state. The initial condition is the only phase-field state
used in training; later reference states serve only for evaluation. The training signal therefore
comes from the governing residual on the model rollout, not from reference trajectories as in
trajectory-supervised surrogates. In contrast to phase-wise coordinate approaches such as PINNs-MPF,
all channels advance recurrently with one shared field model. A complementary phase-local realization (Section~\ref{sec:methods:phaselocal}) keeps
this physics-trained, autonomous formulation but uses a shared finite-neighborhood rule and a
projected-obstacle reference, so its 2,160 learned parameters do not depend on the number of grains.
The two realizations are trained separately and are not ranked against each other. This paper
supersedes the preliminary study~\cite{elfetni2025preprint}.

\Needspace{6\baselineskip}
Figure~\ref{fig:framework} summarizes both realizations. Section~\ref{sec:methods} defines the
operators, admissibility maps, training objectives and evaluation protocols.
Section~\ref{sec:results} follows the four parts. Sections~\ref{sec:results:architecture}
to~\ref{sec:results:specialization} cover foundation and generalization, and
Sections~\ref{sec:results:phaselocal} to~\ref{sec:results:population} transfer across domain size and
grain number. Sections~\ref{sec:results:fidelity} and~\ref{sec:results:controls} address
grain-resolved fidelity, and Section~\ref{sec:results:screening} the virtual hold. Section~\ref{sec:discussion} discusses the
contribution, related work and scope. All comparisons assess numerical fidelity to the specified
discrete reference in normalized simulation time, not material calibration.

\skelfig{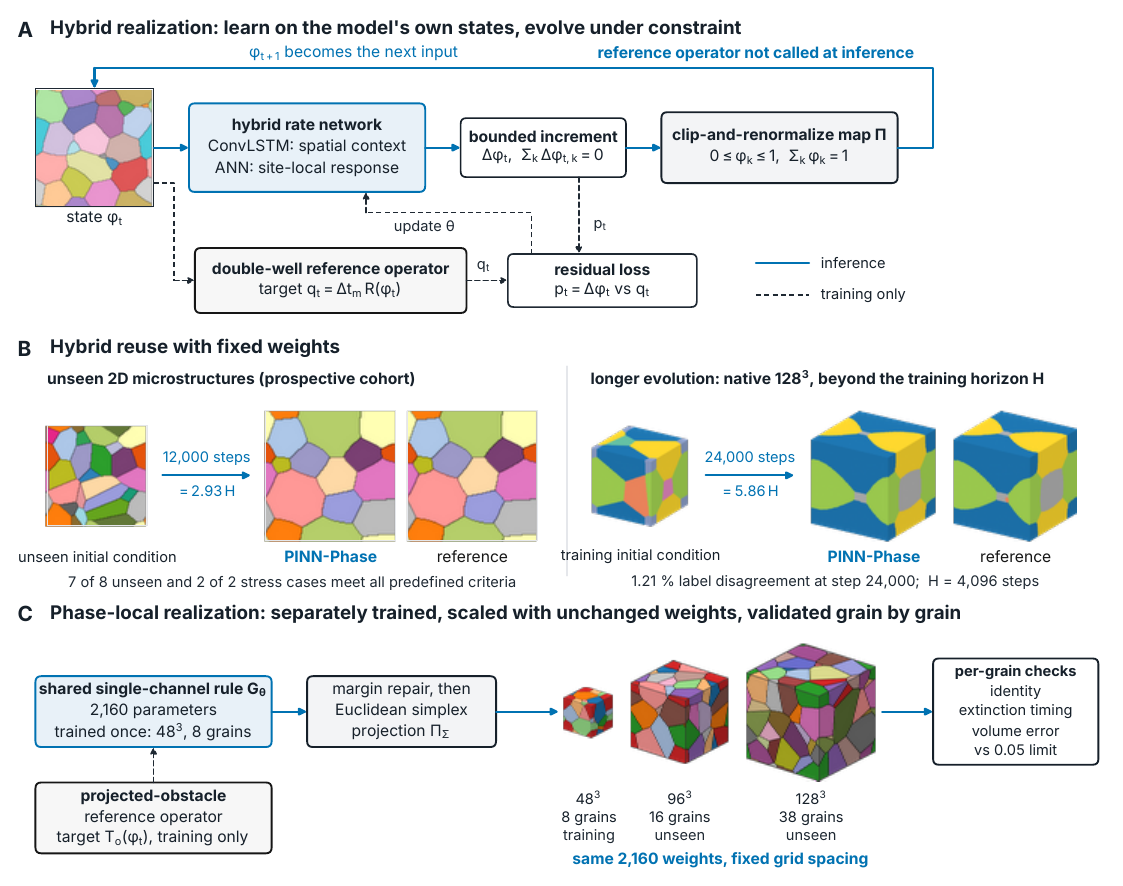}{%
\footnotesize\textbf{Two complementary realizations of PINN-Phase.}
(a)~Hybrid realization (ConvLSTM: convolutional long short-term memory; ANN: site-local neural
network). In training (dashed arrows), the double-well reference operator $\mathbf{R}$ evaluated on
the model's own state gives the target $\mathbf{q}_t$ of each learned increment; in inference (solid
arrows), the bounded increment and the clip-and-renormalize map $\Pi$ advance the field without
calling the reference. (b)~The same hybrid weights applied to an unseen two-dimensional
microstructure for 12,000 steps ($2.93H$; Figure~\ref{fig:n25gen}) and to the native $128^3$
training case for 24,000 steps ($5.86H$; Figure~\ref{fig:n128cube}), beside the reference;
$H=4{,}096$ steps is the training horizon. (c)~The separately trained phase-local realization: one
shared 2,160-parameter rule with margin repair and simplex projection $\Pi_\Sigma$, trained at $48^3$
with eight grains and applied unchanged to larger domains and grain numbers
(Figures~\ref{fig:phaselocal}, \ref{fig:population} and~\ref{fig:fidelity}). Colors denote grain
identity.}{fig:framework}
\end{revision}

\begin{revision}
\section{Method}
\label{sec:methods}

\subsection{Field representation and problem scope}
\label{sec:methods:problem}

For the explicit multiphase problems, the state is an $N$-component diffuse field
$\boldsymbol{\varphi}(\mathbf{x},t)=(\varphi_1,\ldots,\varphi_N)$ defined on a periodic two- or
three-dimensional grid. Each component is a local phase fraction and the field is admissible when
\begin{equation}
\mathcal{A}_N=
\left\{\boldsymbol{\varphi}\in[0,1]^N:\sum_{k=1}^{N}\varphi_k=1\right\}.
\label{eq:admissible}
\end{equation}
The scalar benchmarks use one order parameter in an implicit matrix, rather than the $N=1$ limit
of Equation~\eqref{eq:admissible}, which would be identically one.

All simulations are periodic and use the normalized grid units of the reference calculations
(Supplementary Table~\ref{tab:s_physparams}). Table~\ref{tab:matrix} summarizes the benchmarks;
Section~\ref{sec:methods:phaselocal} defines the separate phase-local realization.

\subsection{Projected multiphase-field operator}
\label{sec:methods:physics}

The explicit MPF benchmarks use an equal-parameter, isotropic projected multichannel Allen--Cahn
operator~\cite{allen1979microscopic,garcke1999multiphase,nestler2008preserved,wu2017multiphase}.
For a vector
$\mathbf{v}\in\mathbb{R}^N$, define the zero-sum projection
\begin{equation}
\mathcal{P}_N(\mathbf{v})_k
= v_k-\frac{1}{N}\sum_{j=1}^{N}v_j,
\qquad
\sum_{k=1}^{N}\mathcal{P}_N(\mathbf{v})_k=0.
\label{eq:proj}
\end{equation}
Let
\begin{equation}
W(\varphi)=\varphi^2(1-\varphi)^2,
\qquad
W'(\varphi)=2\varphi(1-\varphi)(1-2\varphi).
\label{eq:doublewell}
\end{equation}
The rate evaluated in training and used by the direct reference integrator is
\begin{equation}
\mathbf{R}(\boldsymbol{\varphi})
=\mu\sigma\,
\mathcal{P}_N\!\left[
\nabla^2\boldsymbol{\varphi}
-\frac{1}{\eta^2}W'(\boldsymbol{\varphi})
\right],
\label{eq:residual}
\end{equation}
where $\mu$ is the interface mobility, $\sigma$ the interfacial energy scale, $\eta$ the interface
width, and $W'$ acts componentwise. The zero-sum rate is curvature-driven; anisotropic pairwise
energies, heterogeneous mobilities and bulk driving forces are outside this operator. It is related
to the framework of Steinbach et al.~\cite{steinbach1996phase}, but references are direct
integrations of Equation~\eqref{eq:residual}, not outputs of an independent Steinbach-type solver.

On a grid $\Omega_h$ with spacing $\Delta x$, the Laplacian is evaluated by second-order central
differences,
\begin{equation}
(\nabla_h^2\boldsymbol{\varphi})_{\mathbf{i}}
=\frac{1}{\Delta x^2}\sum_{m=1}^{d}
\left(\boldsymbol{\varphi}_{\mathbf{i}+\mathbf{e}_m}
-2\boldsymbol{\varphi}_{\mathbf{i}}
+\boldsymbol{\varphi}_{\mathbf{i}-\mathbf{e}_m}\right),
\label{eq:laplacian}
\end{equation}
with periodic indexing in all $d$ directions.

The scalar benchmarks follow the double-obstacle-type Allen--Cahn form of the scalar reference
calculations,
\begin{equation}
\frac{\partial\varphi}{\partial t}
=\mu\left[
\sigma\left(\nabla^2\varphi+\frac{\pi^2}{2\eta^2}(2\varphi-1)\right)
+\frac{\pi}{\eta}\sqrt{\varphi(1-\varphi)}\,\Delta g
\right],
\label{eq:scalarac}
\end{equation}
where $\Delta g=0$ in the curvature-driven studies. The different local potentials preclude treating
this as the one-channel limit of Equation~\eqref{eq:residual}.

In the reported scalar runs, the learned derivative and its physical target both pass through a
bound-aware rate map before the update. Negative outward rates are zeroed where
\(\varphi\le10^{-3}\), and positive outward rates are zeroed where \(\varphi\ge1-10^{-3}\).
The state is then clipped to \([0,1]\). The scalar PDE loss thus compares the learned derivative with the effective
bound-aware Allen--Cahn rate, while Equation~\eqref{eq:scalarac} is the unconstrained operator of
the direct reference step before clipping.

For curvature-driven shrinkage ($\Delta g=0$), the sharp-interface area law for an isolated
circular grain is
\begin{equation}
R^2(t)=R_0^2-2\mu\sigma t,
\qquad
t_{\mathrm{ext}}=\frac{R_0^2}{2\mu\sigma},
\label{eq:radiuslaw}
\end{equation}
where $R_0$ is the initial radius~\cite{mullins1956twodimensional}. This is a benchmark observable,
not an inferred asymptotic grain-growth law.

\subsection{Hybrid neural rate operator}
\label{sec:methods:hybrid}

PINN-Phase predicts its update from the current model field by a learned convex blend of two
complementary branch responses,
\begin{equation}
\mathbf{z}_t=f_\theta(\boldsymbol{\varphi}_t)
=\gamma\,\mathbf{z}^{\mathrm{loc}}_t+(1-\gamma)\,\mathbf{z}^{\mathrm{sp}}_t,
\qquad \gamma\in(0,1),
\label{eq:ratesum}
\end{equation}
where $\gamma$ is the sigmoid of one trainable scalar initialized at $1/2$. The shared site-local
network applies to each phase channel; the spatial ConvLSTM branch exchanges neighboring-grid
information and carries recurrent state~\cite{shi2015convlstm}. The local inputs include phase and
coordinate features in the original formulation. The update below bounds their combined response.
Architecture and training settings are in Supplementary Table~\ref{tab:s_training}.

\begin{nthreerev}
The cascade family, two-dimensional prospective cohort and dense $N=64$ studies use a distinct
permutation-equivariant formulation with 9,605 parameters. Phase channels are interchangeable and
periodic translations preserve the physics: the site-local branch therefore receives only the
local fraction, channel maps are shared with symmetric cross-phase aggregation, and the spatial
branch uses circular padding~\cite{zaheer2017deepsets,cohen2016gcnn}. The global blend remains;
there is no graph conditioner. The prospective $N=16$, $96^3$ cohort of
Section~\ref{sec:methods:n16cohort} instead retains the original development model and scores raw
phase identities without relabeling.
\end{nthreerev}

\subsection{Bounded time integration and structural admissibility}
\label{sec:methods:admissibility}

With effective model integration interval $\Delta t_{\mathrm{m}}$ and fixed increment scale
$\Delta\varphi_{\max}$, the combined neural response is squashed by a hyperbolic tangent and
scaled. The zero-sum projection of the physical operator then centers it across phase channels:
\begin{equation}
\Delta\boldsymbol{\varphi}_t
=\Delta\varphi_{\max}\,
\mathcal{P}_N\!\left[\tanh(\mathbf{z}_t)\right],
\qquad
\boldsymbol{\psi}_{t+1}
=\boldsymbol{\varphi}_t+\Delta\boldsymbol{\varphi}_t.
\label{eq:bounded}
\end{equation}
The bounded increment is phase-sum neutral. Its associated learned rate is
$\mathbf{r}_t=\Delta\boldsymbol{\varphi}_t/\Delta t_{\mathrm{m}}$, with
$\Delta t_{\mathrm{m}}=\Delta t_{\mu\sigma}/(\mu\sigma)$ for explicit MPF
(Supplementary Table~\ref{tab:s_symbolmap}).

Admissibility is imposed after every update. For a raw phase vector $\boldsymbol{\psi}$, define
\begin{equation}
\bar\psi_k=\min\!\left(1,\max(0,\psi_k)\right),
\qquad
s=\sum_{j=1}^{N}\bar\psi_j,
\label{eq:clip}
\end{equation}
and
\begin{equation}
\Pi_k(\boldsymbol{\psi})=
\begin{cases}
\bar\psi_k/s, & s>\varepsilon_\Pi,\\[2mm]
1/N, & s\le\varepsilon_\Pi,
\end{cases}
\qquad
\boldsymbol{\varphi}_{t+1}=\Pi(\boldsymbol{\psi}_{t+1}).
\label{eq:admissibility_map}
\end{equation}
Here $\varepsilon_\Pi>0$ guards the denominator without removing state components. This
clip-and-renormalize admissibility map guarantees bounds and unit sum; it is not a Euclidean
nearest-point simplex projection. Optional sub-threshold pruning is a separate map variant
(Supplementary Table~\ref{tab:s_training}). The final $N=16$, $96^3$ continuation uses the strict
map in training and prediction.

Figure~\ref{fig:admissibility} distinguishes the physical-rate projection, bounded increment and
post-update map; only the map enforces admissibility of the carried state. A pre-map phase-sum
penalty, where enabled, is auxiliary.

\skelfig{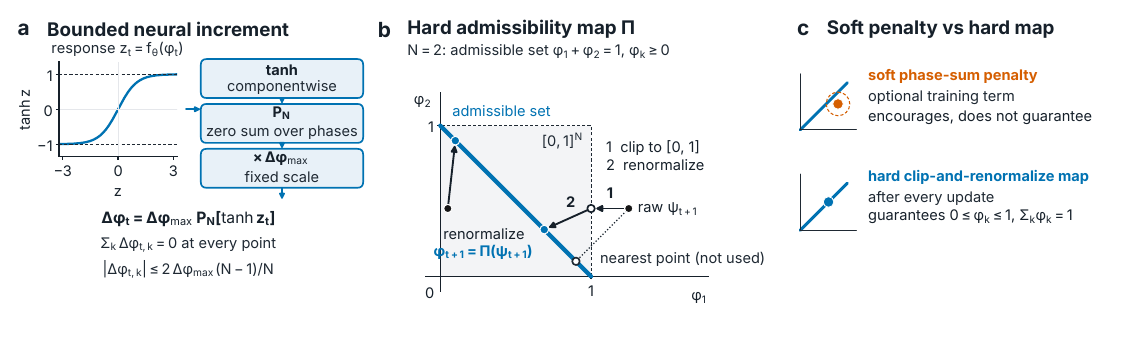}{%
\textbf{Bounded neural update and hard multiphase admissibility.} (a)~The response
$\mathbf{z}_t$ passes through a componentwise hyperbolic tangent, the zero-sum projection
$\mathcal{P}_N$, and the scale $\Delta\varphi_{\max}$. The resulting increment is bounded and
phase-sum neutral at every grid point. (b)~The map $\Pi$ is shown for $N=2$, where the admissible
set is $\varphi_1+\varphi_2=1$ with $\varphi_k\ge0$. The raw vector
$\boldsymbol{\psi}_{t+1}$ is clipped to $[0,1]$ and renormalized to unit sum along its ray through
the origin. The hollow marker shows the Euclidean nearest-point projection, which is not used.
(c)~The optional soft phase-sum penalty encourages admissibility during training but cannot
guarantee it. The hard map enforces bounds and unit sum after every update.}{fig:admissibility}

\subsection{Physics-informed training objective}
\label{sec:methods:training}

Training starts from the initial field $\boldsymbol{\varphi}_0$. At each neural step, the physical
target increment is evaluated on the current predicted field. It is compared with the bounded
increment of Equation~\eqref{eq:bounded}:
\begin{equation}
\mathbf{q}_t
=\Delta t_{\mathrm{m}}\,\mathbf{R}(\boldsymbol{\varphi}_t),
\qquad
\mathbf{p}_t
=\Delta\boldsymbol{\varphi}_t.
\label{eq:target_increment}
\end{equation}
The per-step loss combines relative and absolute increment discrepancies,
\begin{equation}
\ell_t
=w_{\mathrm{rel}}\,
\frac{\big\langle(\mathbf{p}_t-\mathbf{q}_t)^2\big\rangle}
{\big\langle\mathbf{q}_t^2\big\rangle+\varepsilon_r}
+w_{\mathrm{abs}}\,
\frac{\big\langle(\mathbf{p}_t-\mathbf{q}_t)^2\big\rangle}
{s_{\mathrm{abs}}^2+\varepsilon_r},
\label{eq:mixedloss}
\end{equation}
where $\langle\cdot\rangle$ averages over channels and grid points, $\varepsilon_r=10^{-8}$ and
$s_{\mathrm{abs}}=0.05$. Weights and variants are listed in Supplementary
Table~\ref{tab:s_training}. Over training horizon $H$,
\begin{equation}
\mathcal{L}_{\mathrm{res}}(\theta)=\frac{1}{H}\sum_{t=0}^{H-1}\ell_t.
\label{eq:objective}
\end{equation}
The complete objective also doubles the first-step residual weight in every explicit-MPF run;
some runs add pre-map phase-sum and convolutional-branch penalties or interface weighting. Scalar
runs use a separate objective with energy and branch-target terms. All definitions and nonzero
weights are in \smsec{b}, Equation~\eqref{eq:s_full_objective} and
Table~\ref{tab:s_objective}. The state advances by Equations~\eqref{eq:bounded}
and~\eqref{eq:admissibility_map}, outside the loss. Truncated backpropagation through time (TBPTT)
detaches recurrent state between windows; the represented horizon and TBPTT length are reported
separately.

Training uses initial fields and physics-computed one-step targets on the model's own states, with
no independently integrated reference evolution trajectories as training examples. No
post-initial reference field enters the loss or checkpoint selection; see \smsec{f}.

\subsection{Optional graph conditioning}
\label{sec:methods:graph}

An optional graph module in the two-dimensional Voronoi studies derives active regions and shared
boundaries from the predicted field, then returns bounded per-node scale and shift factors,
\begin{equation}
\mathbf{z}'
=\boldsymbol{\alpha}(G_t)\odot\mathbf{z}
+\boldsymbol{\beta}(G_t),
\label{eq:graphcond}
\end{equation}
where $G_t=G(\boldsymbol{\varphi}_t)$. The conditioned response enters Equation~\eqref{eq:bounded}.
This module is absent from the three-dimensional benchmarks; details are in SM-B.

\subsection{Reference integration and evaluation metrics}
\label{sec:methods:eval}

The PF references are generated by explicit integration of the same discretized operator used in
the training residual,
\begin{equation}
\boldsymbol{\varphi}^{\mathrm{ref}}_{t+1}
=\Pi\!\left(
\boldsymbol{\varphi}^{\mathrm{ref}}_t
+\Delta t_{\mathrm{m}}\mathbf{R}
(\boldsymbol{\varphi}^{\mathrm{ref}}_t)
\right),
\label{eq:reference_step}
\end{equation}
with the benchmark-specific admissibility-map variant of Supplementary Table~\ref{tab:s_training}.
These comparisons measure learned integration against the same discrete physical model, rather
than against an independent constitutive formulation or experiment.

For the PINN-Phase field $\hat{\boldsymbol{\varphi}}$ and the reference
$\boldsymbol{\varphi}^{\mathrm{ref}}$, the primary categorical metric is the fraction of grid points
with different phase labels,
\begin{equation}
D_t=\frac{1}{|\Omega_h|}\sum_{\mathbf{x}\in\Omega_h}
\mathbf{1}\!\left[
\arg\max_k\hat\varphi_k(\mathbf{x},t)
\ne
\arg\max_k\varphi_k^{\mathrm{ref}}(\mathbf{x},t)
\right].
\label{eq:argmax}
\end{equation}
The corresponding agreement is $1-D_t$. Field fidelity is measured by the mean absolute error
\begin{equation}
\mathrm{MAE}_t
=\frac{1}{N|\Omega_h|}
\sum_{k=1}^{N}\sum_{\mathbf{x}\in\Omega_h}
\left|\hat\varphi_k(\mathbf{x},t)-
\varphi_k^{\mathrm{ref}}(\mathbf{x},t)\right|.
\label{eq:mae}
\end{equation}
The unchanged-initial-state baseline holds the initial field fixed at every comparison time.

For the explicit MPF benchmarks, the active phase set is defined from the argmax partition,
\begin{equation}
\mathcal{S}_t=
\left\{k:\exists\,\mathbf{x}\in\Omega_h\ \text{such that}\ 
k=\arg\max_j\varphi_j(\mathbf{x},t)\right\}.
\label{eq:active_set}
\end{equation}
A phase extinction is recorded at the first saved frame at which its label is absent from
$\mathcal{S}_t$ and stays absent. Event times are resolved only at the recorded cadence.
The native $128^3$ study uses separate 20-step identity records (Supplementary Section~\ref{sm:j}).
Scalar connected components, counted after thresholding at $\varphi>1/2$, do not represent
persistent grain identities.

The equivalent radius uses phase-field mass rather than a thresholded region. In two dimensions,
the area is $A=\sum_{\mathbf{x}\in\Omega_h}\varphi(\mathbf{x})\,\Delta x\,\Delta y$; in three,
the volume is $V=\sum_{\mathbf{x}\in\Omega_h}\varphi(\mathbf{x})\,\Delta x\,\Delta y\,\Delta z$.
Both are discrete integrals of the order parameter:
\begin{equation}
R_{\mathrm{eq}}^{\mathrm{2D}}=\sqrt{A/\pi},
\qquad
R_{\mathrm{eq}}^{\mathrm{3D}}=\left(\frac{3V}{4\pi}\right)^{1/3}.
\label{eq:radeq}
\end{equation}
The same mass-based estimator is applied to reference and prediction. Scalar coarsening uses the
interface-band proxy
\begin{equation}
L_t=
\left|\left\{\mathbf{x}\in\Omega_h:
0.1\le\varphi(\mathbf{x},t)\le0.9\right\}\right|.
\label{eq:ifaceproxy}
\end{equation}
This grid-dependent count is not a calibrated interfacial length. Supplementary Section~\smsec{c}
defines the normalized energy and additional component statistics.

\subsection{Benchmark design and implementation}
\label{sec:methods:benchmarks}

The benchmarks progress from interface motion and coalescence to multiphase relaxation, long-time
grain extinction, and three-dimensional evolution. The four-phase junction probes sensitivity to
the represented training horizon. The $N=8$ and $N=16$ cubes test one and three designed grain
extinctions, respectively. Table~\ref{tab:matrix} lists each domain, phase count, training horizon,
evaluation quantity, and scientific role.

The hybrid models retain the two-branch local--spatial design selected in the scalar ablation.
\mrrev{The cascade-family, prospective two-dimensional cohort, and dense $N=64$ studies use the
permutation-equivariant formulation of Section~\ref{sec:methods:hybrid}.} Capacity and sequence
settings vary by benchmark; Supplementary Table~\ref{tab:s_training} gives the training settings
and admissibility maps.
\end{revision}

\begin{genrev}
\subsection{Prospective evaluation on unseen \nsixteencell{two-dimensional} initial conditions}
\label{sec:methods:generalisation}

The hybrid model was tested on eight previously unseen two-dimensional microstructures and two
predefined stress cases with unusually dense or sparse late-stage grain populations. Cases,
references, scoring criteria, and model checkpoints were fixed before evaluation. The primary
model had been trained on one development initial condition and was rolled out with unchanged
weights for 12{,}000 steps. A secondary model trained on six initial conditions was descriptive;
neither model used graph conditioning (Supplementary Section~\smsec{g}).

A case passed only if its field, interface, grain-identity, and numerical-integrity criteria all
held. In particular, no reference survivor could be missing from the prediction. The cohort
required at least seven of eight unseen cases and both stress cases to pass, alongside its
predefined aggregate gain and identity conditions. Extinction timing was descriptive and did not
affect qualification. Table~\ref{tab:n25gen} reports the outcomes, while Supplementary
Table~\ref{tab:s_n25gen_criteria} gives every threshold, observed range, and the required
challenge to the unchanged-initial-state baseline.

The cohort was scored once against independently generated references, without revising cases,
thresholds, or model weights. Predictions ended at step 12{,}000. A reference-only continuation
to step 13{,}000 provides descriptive context and does not alter any qualification
(Supplementary Section~\smsec{g}).

\subsection{Long-horizon evaluation at 64 grains}
\label{sec:methods:n64}

\crev{The dense 64-grain benchmark compares a single predefined primary evaluation with a later
training continuation. The primary graph-free, permutation-equivariant model was advanced for
12{,}000 steps from its development initial condition and scored against the matching direct
phase-field reference. Its checkpoint and acceptance criteria were fixed before evaluation.}
\end{genrev}

\begin{e1rev}
\crev{After primary scoring, training resumed from the same checkpoint with a longer represented
horizon, followed by a second 12{,}000-step comparison. The continuation retained the network,
loss, admissibility map, initial condition, reference, and scoring rule. It changed both the
represented horizon and training effort, so it tests their joint sensitivity rather than an
isolated horizon effect or an independent evaluation. Supplementary Table~\ref{tab:s_n64_horizon}
gives both outcomes and the full training and saved-state settings.}
\end{e1rev}

\begin{n16rev}
\subsection{Prospective three-dimensional cohort evaluation}
\label{sec:methods:n16cohort}

One fixed hybrid model was evaluated on six prospectively selected $N=16$, $96^3$ initial
microstructures from the development cube's designed-extinction family. Only grain arrangement
varied; the physical operator, discretization, and three-extinction design were unchanged. No
case informed training, tuning, checkpoint choice, or adaptation. Model rollouts were completed
before later reference states were generated or opened for scoring.

Raw grain labels were compared without alignment or relabeling. The scored state is step 1600,
after the designed extinctions; states through step 3200 test whether the correct survivor set
persists. A case passed only if all six predefined conditions held: numerical integrity, terminal
field agreement, improvement over the unchanged initial state, exact survivor identities,
extinction identity and wave order, and tail survivor-set agreement. Individual event times are
resolved at the 200-step saved-state cadence. Supplementary Section~\ref{sm:i} gives the exact
thresholds, gain definition, and per-case scores.
\end{n16rev}

\subsection{Native \texorpdfstring{$128^3$}{128\textthreesuperior} multi-initial-condition training}
\label{sec:methods:n128}

The native-grid model learns directly on eight phase fields over $128^3$ voxels, without coarse-grid prediction or upsampling. Its training residual and the direct phase-field reference use the same projected multiphase-field operator and clip-and-renormalize admissibility map. The local network and convolutional recurrent branch are retained, while graph conditioning is disabled. This model starts from the common earlier initialization, independently of the single-initial-condition predecessor's final weights (Supplementary Section~\ref{sm:j}).

Four initial fields evolve separately during training, and each update combines their physics-residual gradients. The fields include the preceding native-grid benchmark and three additional draws from the same initial-condition family. No post-initial reference field enters the loss, state updates or checkpoint selection. The final epoch is selected by position; evaluation on a training field checks retained fidelity, not independent transfer. Architecture, objective, optimizer history and training-field handling are specified in Supplementary Section~\ref{sm:k} and Table~\ref{tab:s_multiic_training}.

\subsection{Native \texorpdfstring{$128^3$}{128\textthreesuperior} transfer and long-horizon evaluation}
\label{sec:methods:n128transfer}

Unchanged weights are evaluated on two separate cohorts of six initial conditions absent from training. The development references had been examined previously; the unseen cohort and its scoring procedure were fixed before those outcomes were opened. Case numbers are local to each cohort, and their results are never pooled.

Each autonomous prediction and paired direct phase-field reference run from its own initial field to step $24{,}000$, or $5.86$ times the represented training horizon. Fixed initial phase identities determine grain correspondence without registration or relabeling. Recorded grain disappearance is assessed at the ownership-count cadence, so timing residuals have interval uncertainty. Full field and event definitions appear in Supplementary Section~\ref{sm:l}.

Passing requires agreement in final labels, improvement over the unchanged initial state, terminal active-grain count and survivor identities, extinction count, order and timing, and the fate of a designated small grain. The predefined cohort rule, exact thresholds and case-level outcomes are in Supplementary Section~\ref{sm:l}. The within-horizon saved states provide descriptive context and do not replace full-horizon scoring.

\subsection{Case-specific physics-informed continuation}
\label{sec:methods:specialization}

Three previously inspected development microstructures test whether further physics-informed training can repair a missed long-horizon criterion while retaining earlier successes. Two had missed extinction timing; the third had satisfied all six criteria. Each continuation starts independently from the multi-initial-condition parent and uses only its selected initial field and the unchanged physics objective. The adapted prediction is then evaluated to step $24{,}000$ under the original rule. The update budget, optimizer settings and reference isolation are documented in Supplementary Section~\ref{sm:m} and Table~\ref{tab:s_multiic_training}.

Success in the two failing cases requires bringing the final extinction residual within the predefined $\pm800$-step band without losing another satisfied criterion. The passing case must retain all six criteria. These requirements were fixed before each continuation, after the direct-transfer outcomes were known. This retrospective study does not establish a reference-independent adaptation trigger or a comparison with training from a random initialization. Per-case outcomes and phase-volume histories are reported in Supplementary Section~\ref{sm:m}.

\subsection{Phase-local realization}
\label{sec:methods:phaselocal}

\emph{Common form and design requirement.}
Both realizations advance the carried state by a learned increment followed by a feasibility map,
and both obtain their physical training target on the model's own generated states:
\begin{equation}
\boldsymbol{\varphi}_{t+1}
=\mathcal{A}_{\mathcal{F}}\!\left(\boldsymbol{\varphi}_t,\;\boldsymbol{\varphi}_t+\Delta_\theta(\boldsymbol{\varphi}_t)\right).
\label{eq:commonform}
\end{equation}
Equation~\eqref{eq:commonform} is a conceptual common form, not an identity of the two
realizations. In the hybrid realization, $\Delta_\theta$ is the bounded, phase-sum-neutral increment
of Equation~\eqref{eq:bounded}, which also depends on the recurrent hidden state, and
$\mathcal{A}_{\mathcal{F}}(\boldsymbol{\varphi},\boldsymbol{\psi})=\Pi(\boldsymbol{\psi})$ is the
clip-and-renormalize map of Equation~\eqref{eq:admissibility_map}. In the phase-local realization,
$\Delta_\theta$ applies one shared single-channel rule to each phase, and $\mathcal{A}_{\mathcal{F}}$
is the margin repair and exact simplex projection of Equation~\eqref{eq:localstep}, which also reads
the carried state. This realization meets a different requirement: the number of learned parameters
must not depend on the phase count $N$. The hybrid convolutional-recurrent branch has input and output
weights dimensioned by $N$, and its permutation-equivariant variant (Section~\ref{sec:methods:hybrid})
was trained and evaluated at a fixed phase count. In the phase-local realization, every phase channel
passes through the same rule without learned cross-phase aggregation, and channels interact only
through the constraint map, which couples all phases at every voxel. Its reference operator,
constraint map and loss also differ; it is separately trained, not a simplified hybrid checkpoint.

\emph{Shared local rule and constraint map.}
Let $u$ denote one phase field on the periodic grid. The shared rule is
\begin{equation}
G_\theta(u)=W_3\ast\tanh\!\left(W_2\ast\tanh\!\left(W_1\ast u\right)\right),
\label{eq:glocal}
\end{equation}
where $W_1$, $W_2$ and $W_3$ are bias-free $3\times3\times3$ convolutions with channel widths
$1\rightarrow8$, $8\rightarrow8$ and $8\rightarrow1$ and periodic padding. The rule has
$27(8+64+8)=2{,}160$ learned parameters, a receptive radius of three voxels, no recurrent hidden
state, and no phase-index or coordinate input. At every voxel the learned step is
\begin{equation}
\begin{split}
z_k&=\varphi_k+G_\theta(\varphi_k),\\
\tilde z_k&=z_k+\varphi_k\max\!\Big(0,\;1+\Delta t_{\mathrm{o}}\beta_{\mathrm{o}}-\sum_{j=1}^{N}z_j\Big),\\
\mathcal{U}_\theta(\boldsymbol{\varphi})&=\Pi_{\Sigma}(\tilde{\boldsymbol{z}}),
\end{split}
\label{eq:localstep}
\end{equation}
where $\Pi_\Sigma$ is the exact Euclidean projection onto the phase-fraction simplex
$\Sigma_N=\{\mathbf{v}\in\mathbb{R}^N:v_k\ge0,\ \sum_k v_k=1\}$,
\begin{equation}
\left[\Pi_{\Sigma}(\mathbf{v})\right]_k=\max\!\left(v_k-\tau(\mathbf{v}),0\right),
\qquad
\sum_{k=1}^{N}\max\!\left(v_k-\tau(\mathbf{v}),0\right)=1,
\label{eq:simplexproj}
\end{equation}
with one threshold $\tau(\mathbf{v})$ per voxel, found by sorting the
components~\cite{condat2016simplex}. Unlike the hybrid map $\Pi$, $\Pi_\Sigma$ returns the nearest
admissible point. The margin repair raises the pre-projection phase sum to at least
$1+\Delta t_{\mathrm{o}}\beta_{\mathrm{o}}$, which, with the bias-free network, keeps a phase that is
absent everywhere exactly zero. One learned step represents one reference step, and no reference
operator is called during learned prediction.

\emph{Projected obstacle reference map.}
The reference for this realization is a different discrete operator from the double-well family of
Equations~\eqref{eq:residual} and~\eqref{eq:reference_step}, which remains the hybrid reference. It is
a projected obstacle-type multiphase-field map with a phase-fraction simplex
constraint~\cite{garcke1999multiphase},
\begin{equation}
\mathcal{T}_{\mathrm{o}}(\boldsymbol{\varphi})
=\Pi_{\Sigma}\!\left[\boldsymbol{\varphi}
+\Delta t_{\mathrm{o}}\left(\kappa_{\mathrm{o}}\nabla_h^2\boldsymbol{\varphi}
+\beta_{\mathrm{o}}\boldsymbol{\varphi}\right)\right],
\label{eq:obstacleref}
\end{equation}
where $\nabla_h^2$ is the seven-point periodic Laplacian of Equation~\eqref{eq:laplacian},
$\Delta x=1$, $\Delta t_{\mathrm{o}}=0.01$, $\kappa_{\mathrm{o}}=64/\pi^2\approx6.485$ and
$\beta_{\mathrm{o}}=1$; the diffuse interface width $\pi\sqrt{\kappa_{\mathrm{o}}/\beta_{\mathrm{o}}}$ is
eight grid spacings. The operator and constraint map are part of the tested formulation, so the
results are attributed to the complete learned update rather than to the network alone. The target
is one step of a linear nearest-neighbor stencil per phase plus the exact projection; locality and
transfer across domain size and grain count are expected properties of this target. The experiments
test whether a learned nonlinear local map, trained on small domains, reproduces it stably over long
autonomous rollouts under the stated criteria.

\emph{Training.}
One model was trained once on four initial microstructures at $48^3$ with eight grains, with 28{,}200
Adam updates. Each update samples six $20^3$ crops (with a three-voxel halo) from a buffer that holds
the four initial fields and states of the model's own autonomous rollouts. The target of each crop
is one application of $\mathcal{T}_{\mathrm{o}}$ to the sampled state, detached from the graph; it
does not advance any trajectory. The loss is
\begin{equation}
\mathcal{L}_{\mathrm{local}}(\theta)
=\Big\langle\sum_{k=1}^{N}
\big[\mathcal{U}_\theta(\boldsymbol{\varphi})_k
-\operatorname{sg}\!\big(\mathcal{T}_{\mathrm{o}}(\boldsymbol{\varphi})\big)_k\big]^2\Big\rangle_{\text{crops, voxels}},
\label{eq:localloss}
\end{equation}
with $\operatorname{sg}$ the stop-gradient. No independently evolved reference trajectory supplies a
training state or target (schedule, buffer and training record in Supplementary
Section~\ref{sm:phaselocal}).

\emph{Evaluation regimes.}
The checkpoint is evaluated unchanged, without retraining, tuning or case-specific parameters, in
three regimes at fixed grid spacing. The $64^3$ regime with 16 grains runs over 3{,}000 steps
($t=30$); the $96^3$ regime with 16 grains and the $128^3$ regime with 38 grains run over 6{,}750 steps
($t=67.5$). The longer horizon scales
3{,}000 steps by the parabolic-coarsening factor $2.25$ of the mean initial grain volume and was fixed
before any case was drawn. Each regime uses three unseen periodic Voronoi microstructures of one
synthetic geometry family, admitted by a rule fixed in advance from reference evolution alone: the
first generated microstructures whose reference evolution contains at least three extinctions and no
return of an extinct phase. Regimes are reported separately and never pooled; the $96^3$ evaluation
served as a decision check before the $128^3$ microstructures were generated.

\emph{Acceptance criteria.}
Let $V_k(s)=\sum_{\mathbf{x}}\varphi_k(\mathbf{x},s)$ be the volume of phase $k$ after step $s$. Phase
$k$ is alive when $V_k(s)>10^{-6}$, and its extinction step is the first $s\ge1$ with
$V_k(s)\le10^{-6}$. A case is accepted only if all of the following hold, with the time-dependent
conditions assessed at the eight ($64^3$) or twelve (larger regimes) evaluation times. (i) The
structural permutation and zero-phase-padding checks pass. (ii) Every step is completed with finite,
nonnegative fields and a unit-sum error of at most $10^{-12}$. (iii) The alive and extinct sets equal
the reference's, each extinction step lies within $\max(100,\,0.1\,s_{\mathrm{ref}})$ steps of the
reference step $s_{\mathrm{ref}}$, and no extinct phase returns. (iv) The continuous-field error
$E=\sum_{\mathbf{x},k}|\varphi^{\mathrm{model}}_k-\varphi^{\mathrm{ref}}_k|/(2n^3)$ on the $n^3$ grid is
at most half that of static persistence at every evaluation time from $t=5$, and at most one quarter
at the terminal time. (v) The maximum normalized grain-volume error
$\max_k|V^{\mathrm{model}}_k-V^{\mathrm{ref}}_k|/(n^3/N)$ is at most $0.05$. Extinction identity and
timing come from the recorded volumes, not from argmax labels; exact definitions and evaluation times
are in Supplementary Section~\ref{sm:phaselocal}.

\emph{Matched comparator.}
A matched comparator has the same architecture, loss and update budget but is trained on
reference-trajectory states: its buffer is filled by advancing the four initial fields with the
reference map. It tests whether training on the model's own states is needed for the transfer. It
was evaluated at $96^3$ and $128^3$, not at $64^3$, and never determines the primary outcome.

\emph{Transfer to 150 grains.}
The unchanged checkpoint was then applied at $128^3$ with 150 grains, whose mean initial grain volume
($13{,}981$ voxels) is comparable to that of training ($13{,}824$); the grain-size distribution is not
controlled. The 13{,}500-step horizon ($t=135$) and the first three qualifying microstructures, under
the same admission rule with all phases initially present, were fixed before any model run. The
criteria are unchanged and assessed at twelve evaluation times (Supplementary
Section~\ref{sm:n150}).

\subsection{Evaluation with thousands of grains and decision-level analysis}
\label{sec:methods:population}

\emph{Grain remapping.}
For the 150-, 1,000- and 5,000-grain evaluations, the learned model and the reference run on a
remapped representation in which one computational channel can carry several grains, each keeping its
own identity. Two grains share a channel only when their supports are separated by more than $2r$,
where $r=3$ voxels is the receptive radius of $G_\theta$, and channels are reassigned before any pair
can violate the separation. Because a phase vanishes outside its interface and inactive channels
remain zero, Equations~\eqref{eq:localstep} and~\eqref{eq:obstacleref} act on the remapped field as on
the dense field in exact arithmetic; the remapping changes the storage, not the weights or the grain
identities. At $128^3$ with 150 grains, remapped and dense runs agree over 13,500 steps to
$4\times10^{-8}$ of the mean initial grain volume, with identical extinction steps. The number of
channels in use stays between 21 and 37 for the learned model and between 14 and 23 for the reference.

\emph{Transfer to 1,000 and 5,000 grains.}
The unchanged weights were applied at $240^3$ with 1,000 grains (three microstructures, 13,500 steps,
$t=135$) and at $416^3$ with 5,000 grains (one microstructure, 5,000 steps, $t=50$), from the same
periodic Voronoi generator. The mean initial grain volumes, $13{,}824$ and $14{,}398$ voxels, are
comparable to the 150-grain set, so the grain count grows at a nearly constant mean grain size. The
5,000-grain horizon was fixed before any of its results were available. The criteria of the 150-grain
regime apply unchanged. In addition, the surviving fraction $N(t)/N_0$, the mean grain size
$\bar R^2$ of Section~\ref{sec:results:n150} and the per-grain volume error
$|V^{\mathrm{model}}_k-V^{\mathrm{ref}}_k|/\bar V_0$ over grains alive in both runs are compared, with
$\bar V_0$ the mean initial grain volume. Each run used one NVIDIA H100 GPU; run times of both methods
are listed in the Supplementary Material.

\emph{State-matched statistics and time-rescaling controls.}
Two controls replay the reference on a uniformly faster clock: the control state at normalized time
$t$ is the reference state at $m\,t$, with $m=1.03$ or $1.10$. This is the trajectory that a uniform
3\% or 10\% increase of the boundary mobility would produce. Grain volumes are interpolated linearly between saved
steps and never extrapolated, so the controls are defined up to $t=50/m$ ($48.54$ and $45.45$) at
5,000 grains. A uniform time rescaling leaves every statistic indexed by the state unchanged. Two are compared:
the mean number of faces per grain as a function of $\bar R^2$, and the distribution of normalized
radii $R/\langle R\rangle$ at $N/N_0=0.60$. The latter is summarized by the Kolmogorov--Smirnov
distance and by the exponent $\gamma$ of the generalized Hillert distribution with the same ratio
$\langle R\rangle^2/\langle R^2\rangle$~\cite{hillert1965}. It changes $N(t)$, $\bar R^2(t)$,
per-grain volumes at a fixed time and extinction times. State-matched statistics therefore cannot
separate the controls from the reference by construction, whereas time-indexed and grain-resolved
comparisons can.

\emph{Virtual isothermal hold.}
The virtual hold is a numerical analogue of choosing how long to hold a polycrystal at a fixed
temperature. The coarsening measure $Q(t)=M(t)/M_0$ is evaluated from grain volumes alone, where
$M(t)$ is the mean volume of the alive grains and $M_0$ its initial reference value. Because the total
volume is fixed, $Q$ follows the inverse of the surviving fraction. A target is met at the first time with
$Q\ge Q^\ast$ and a ratio $L\le12$ of the largest to the mean alive volume. For each model, case and
target, the signed relative time-to-target error is scored against the reference and accepted when
its magnitude is strictly below 5\%. A stop time is also chosen on a 0.1-unit grid; a stop is
premature when the reference has not yet met the target at that time. The targets are $Q^\ast=1.01$, $1.02$,
$1.10$, $1.25$ and $1.50$ for the 150- and 1,000-grain runs, which are retrospective. For the
5,000-grain run they are $1.01$, $1.02$, $1.25$ and $1.50$, prospective because rules and targets were
fixed before any output was inspected. This gives 34 case-target decisions per model. Repeated targets on one microstructure are
correlated, so the decisions are not independent tests. No temperature, material or laboratory
schedule is implied (rules and records in Supplementary Section~\ref{sm:screening}).

\begin{revision}
\begin{algorithm}[!tbp]
\centering
\caption{Physics-informed training and autoregressive prediction with PINN-Phase. The physical
residual is evaluated on the predicted field at every step. For the reported explicit multiphase-field
path, post-initial-condition reference fields are not inputs to the training procedure.}
\label{alg:rollout}
\small
\begin{tabular}{@{}r@{\hspace{0.7em}}>{\raggedright\arraybackslash}p{0.84\linewidth}@{}}
\toprule
 & \textbf{Given:} initial field $\boldsymbol{\varphi}_0$; response network $f_\theta$; physical
 residual $\mathbf{R}$; admissibility map $\Pi$; horizon $H$; TBPTT window $B$; effective
 interval $\Delta t_{\mathrm{m}}$; increment scale $\Delta\varphi_{\max}$.\\[2pt]
1 & \textbf{for} each training epoch \textbf{do} \\
2 & \quad set $\boldsymbol{\varphi}\leftarrow\boldsymbol{\varphi}_0$, initialize the recurrent state, and set the window accumulator $\mathcal{L}_{\mathrm{win}}\leftarrow0$ \\
3 & \quad \textbf{for} $t=0,\ldots,H-1$ \textbf{do} \\
4 & \qquad evaluate the neural response $\mathbf{z}_t=f_\theta(\boldsymbol{\varphi}_t)$ and the model-field target $\mathbf{q}_t=\Delta t_{\mathrm{m}}\mathbf{R}(\boldsymbol{\varphi}_t)$ \\
5 & \qquad compute the bounded increment $\Delta\boldsymbol{\varphi}_t$ from Equation~\eqref{eq:bounded} and add the applicable per-step objective of Equation~\eqref{eq:s_full_objective} to $\mathcal{L}_{\mathrm{win}}$, using $\mathbf{p}_t=\Delta\boldsymbol{\varphi}_t$ and $\mathbf{q}_t$; which of its terms are nonzero in each reported run is given in Table~\ref{tab:s_objective} \\
6 & \qquad set $\boldsymbol{\varphi}_{t+1}=\Pi(\boldsymbol{\varphi}_t+\Delta\boldsymbol{\varphi}_t)$ \\
7 & \qquad \textbf{if} $B$ steps have accumulated \textbf{then} backpropagate $\mathcal{L}_{\mathrm{win}}$, apply the configured gradient clipping, take one optimizer step on $\theta$, zero the gradients, detach both the recurrent state and the carried field $\boldsymbol{\varphi}_{t+1}$, and reset $\mathcal{L}_{\mathrm{win}}\leftarrow0$ \\
8 & \quad \textbf{end for} \\
9 & \textbf{end for} \\
\midrule
 & \textbf{Prediction:} initialize with $\boldsymbol{\varphi}_0$; then at each step evaluate the neural response $\mathbf{z}_t=f_\theta(\boldsymbol{\varphi}_t)$, compute the bounded phase-mean-centered increment $\Delta\boldsymbol{\varphi}_t$ of Equation~\eqref{eq:bounded}, apply the configured admissibility map $\boldsymbol{\varphi}_{t+1}=\Pi(\boldsymbol{\varphi}_t+\Delta\boldsymbol{\varphi}_t)$, and repeat. No physical target is evaluated, no loss is constructed, and no gradients are required.\\
\bottomrule
\end{tabular}
\end{algorithm}
\end{revision}

\par
\begingroup
\footnotesize
\setstretch{1.0}
\setlength{\tabcolsep}{3pt}
\setlength{\LTcapwidth}{\textwidth}
\setlength{\LTpre}{6pt}
\begin{xltabular}{\textwidth}{@{}>{\raggedright\arraybackslash}p{2.3cm} >{\centering\arraybackslash}p{1.15cm} >{\centering\arraybackslash}p{1.0cm} >{\raggedright\arraybackslash}p{3.2cm} >{\raggedright\arraybackslash}p{3.6cm} >{\raggedright\arraybackslash}X@{}}
\caption{Benchmark hierarchy and evaluation design. Physical, numerical and training settings are in Supplementary Tables~\ref{tab:s_physparams} and~\ref{tab:s_training}; the phase-local experiments are specified in Section~\ref{sec:methods:phaselocal} and Supplementary Section~\ref{sm:phaselocal}. \nsixteencell{\textsuperscript{a}Evaluation-only use of the checkpoint of the $N=16$, $96^3$ development cube; no additional training.}}\label{tab:matrix}\\
\toprule
Benchmark & Domain & State size & Training protocol & Primary evaluation & Scientific role \\
\midrule
\endfirsthead
\multicolumn{6}{l}{\tablename\ \thetable\ (continued)}\\
\toprule
Benchmark & Domain & State size & Training protocol & Primary evaluation & Scientific role \\
\midrule
\endhead
\midrule
\endfoot
\bottomrule
\endlastfoot
Scalar shrinkage and two-grain problems & 2D, $128^2$ & one field & case-specific & field MAE, radius law, event timing & interface motion and elementary topology change \\
Scalar multigrain coarsening & 2D, $128^2$ & one field & case-specific & component count, coarsening proxies & descriptive multigrain evolution \\
Four-phase junction & 2D, $128^2$ & $N=4$ & H128--H1024 horizon ladder & label disagreement, junction angles & horizon sensitivity \\
Voronoi coarsening & 2D, $128^2$ & $N=25$ & H4096 & label disagreement to step 12{,}000; active phases & extinction cascade (original hybrid; baseline of Supplementary Figure~\ref{fig:gnn}) \\
Voronoi cascade family & 2D, $128^2$ & $N=25$ & H4096, permutation-equivariant, 9,605 parameters & label disagreement, survivor set, extinction timing & long-horizon coarsening; two development conditions \\
Dense Voronoi coarsening & 2D, $128^2$ & $N=64$ & H4096, permutation-equivariant & label disagreement, survivor set & larger phase count \\
Prospective cohort & 2D, $128^2$ & $N=25$ & evaluation only & complete predefined criterion set & prospective transfer within the cascade family (Table~\ref{tab:n25gen}) \\
Scalar spherical shrinkage & 3D, $64^3$ & one field & fixed scalar-3D model & equivalent-radius trajectory & three-dimensional feasibility \\
Designed-extinction cube & 3D, $64^3$ & $N=8$ & $512\rightarrow2048$ horizon curriculum & label agreement, survivor set, extinction frame & single extinction \\
Designed-extinction cube & 3D, $96^3$ & $N=16$ & H2048 continuation & label agreement, survivor set, three extinction frames & multi-extinction; field limitation in SM-I \\
Prospective $N=16$ cohort & 3D, $96^3$ & $N=16$ & evaluation only\textsuperscript{a} & label disagreement, active set, extinction identity and order & within-family transfer (Table~\ref{tab:n16cohort}) \\
Native $128^3$ cube & 3D, $128^3$ & $N=8$ & $H=4096$; four training initial conditions & label agreement, survivor set, timing to step 24{,}000 & temporal extrapolation to $5.86H$ \\
Development cohort & 3D, $128^3$ & $N=8$ & evaluation only; six cases inspected during development & fields, survivors, events to step 24{,}000 & initial-condition reuse \\
Prospective unseen cohort & 3D, $128^3$ & $N=8$ & evaluation only; fixed before outcomes & label disagreement, survivors, timing to step 24{,}000 & prospective transfer and extrapolation \\
Case-specific continuation & 3D, $128^3$ & $N=8$ & three case-adapted models; 1024 updates & unchanged full-horizon scoring & physics-residual-only adaptation \\
Phase-local realization & 3D, $48^3$ to $128^3$ & $N=8$ to $150$ & 2,160 parameters; own-state one-step loss; 28{,}200 updates & acceptance criteria of Section~\ref{sec:methods:phaselocal} & unchanged-weight transfer across domain size and grain count \\

\end{xltabular}
\endgroup

\begin{revision}
\section{Results}
\label{sec:results}

Table~\ref{tab:headline} summarizes the principal results across the benchmarks in
Table~\ref{tab:matrix}. Field labels use the channel-wise argmax, and $D_t$ denotes label
disagreement as defined in Equation~\eqref{eq:argmax}. The persistence baseline holds the
initial field unchanged; figures label it static-$t_0$.
\par
\begingroup
\footnotesize
\setstretch{1.0}
\setlength{\tabcolsep}{3pt}
\setlength{\LTcapwidth}{\textwidth}
\setlength{\LTpre}{6pt}
\begin{xltabular}{\textwidth}{@{}>{\raggedright\arraybackslash}p{3.3cm} >{\raggedright\arraybackslash}p{2.6cm} >{\raggedright\arraybackslash}p{2.4cm} >{\raggedright\arraybackslash}p{2.6cm} >{\raggedright\arraybackslash}X@{}}
\caption{Principal quantitative results, condensed; Supplementary Table~\ref{tab:s_headline_full} gives every per-row value. Disagreement and MAE are lower-is-better; agreement is higher-is-better. Persistence holds the initial field fixed. The phase-local rows use a different reference operator and acceptance criteria and are not pooled with the hybrid rows.}\label{tab:headline}\\
\toprule
Benchmark & Quantity & PINN-Phase & Reference or baseline & Outcome \\
\midrule
\endfirsthead
\multicolumn{5}{l}{\tablename\ \thetable\ (continued)}\\
\toprule
Benchmark & Quantity & PINN-Phase & Reference or baseline & Outcome \\
\midrule
\endhead
\midrule
\endfoot
\bottomrule
\endlastfoot
Scalar single grain, $128^2$ & full-field MAE & $0.00306$ & $0.03162$ and $0.18944$ (single-branch controls) & radius-law slope within $1.75\%$ \\
Two grains; coalescing grains & full-field MAE & $7.35\times10^{-4}$; $1.430\times10^{-3}$ & PF reference & extinction and coalescence reproduced \\
Four-phase junction & step-6000 label disagreement & $3.272\%$ & $6.543\%$ persistence & junction-angle RMS deviation $5.5^{\circ}$ \\
Voronoi $N=25$, original hybrid & step-12{,}000 label disagreement & $17.20\%$ & $39.70\%$ persistence & extinction-count lag (17 vs.\ 14 phases) \\
Voronoi $N=25$ cascade, two development conditions & step-12{,}000 label disagreement & $1.46\%$; $1.13\%$ & $44.37\%$; $51.25\%$ persistence & exact terminal survivor sets; mostly premature events \\
Dense Voronoi $N=64$ & step-12{,}000 label disagreement & $6.09\%$ & PF reference & all 21 survivors kept, one additional grain \\
Prospective $N=25$ cohort, 2D & cases meeting all criteria & 7 of 8; 2 of 2 stress & criteria fixed in advance & $0.94$--$3.71\%$ against $32.50$--$49.33\%$ persistence \\
Scalar sphere, $64^3$ & radius slope ratio & $0.991$ & PF reference & bound enforcement required \\
MPF cube, $N=8$, $64^3$ & step-2000 label agreement & $99.53\%$ & $94.01\%$ persistence & exact survivor set \\
MPF cube, $N=16$, $96^3$ & step-1600 label agreement & $99.668\%$ & $93.32\%$ persistence & exact survivor set; field limitation in SM-I \\
Prospective $N=16$ cohort, $96^3$ & cases meeting all criteria & 5 of 6 & criteria fixed in advance & exact active set and extinction identities in all six \\
Native $128^3$, training condition & step-24{,}000 label agreement & $98.79\%$ & $69.48\%$ persistence & exact survivor set at $5.86H$ \\
Native $128^3$, development cohort & cases meeting all criteria & four of six & PF reference & all six survivor sets correct \\
Native $128^3$, prospective cohort & step-24{,}000 label disagreement & $2.84$--$11.30\%$ & $27.46$--$33.56\%$ persistence & five of six exact survivor sets \\
Case-specific continuation & last-event residual (steps) & $+1460\rightarrow+520$; $+1340\rightarrow+320$ & $\pm800$-step criterion & mixed volume histories \\
Phase-local, $64^3$/16, $96^3$/16, $128^3$/38 grains & cases meeting all criteria & 3 of 3 per regime & projected-obstacle reference & all 32 extinctions by identity; volume error at most $0.0110$ \\
Phase-local, $128^3$/150 grains & extinctions by identity; volume criterion & 368 of 368; met to $t=80$ & projected-obstacle reference & criterion met to $t=135$ in 1 of 3 \\
Phase-local, $240^3$/1,000 grains, 3 runs & matched extinctions; volume criterion & 2,431 of 2,432; met to $t=57.5$--$67.5$ & projected-obstacle reference & late errors in large grown grains \\
Phase-local, $416^3$/5,000 grains, 1 run to $t=50$ & matched extinctions; maximum volume error & 2,109 of 2,111; $0.038$ & projected-obstacle reference & label disagreement at most $0.35\%$ \\
Virtual hold, 34 decisions & time-to-target error; premature stops & within $0.29\%$; 1 of 34 & controls $m=1.03$, $1.10$: 34 of 34 & timing tolerance and target attainment differ \\
\end{xltabular}
\endgroup

\subsection{Architecture and structural constraints}
\label{sec:results:architecture}

The hybrid operator follows scalar interface motion more closely than either retained branch
alone (Supplementary Figure~\ref{fig:ablation}). On the $128^2$ benchmark, its full-field MAE is
$0.00306$, versus $0.03162$ for ConvLSTM-only and $0.18944$ for the pointwise ANN-only model.
The same ordering holds at $64^2$, although the training objectives differ there (\smsec{b}).
The pointwise branch has no spatial-neighborhood input and did not reproduce coherent interface
motion in this scalar benchmark; this does not assess the finite-neighborhood phase-local model
of Section~\ref{sec:methods:phaselocal}. The hybrid also tracks the equivalent-radius trajectory,
with a $1.75\%$ slope error at $128^2$. Because branch capacities differ, the comparison does not
isolate architecture from parameter count; the full settings and scores are in \smsec{b}.

Across the explicit MPF benchmarks, predicted fields remain bounded and satisfy the unit-sum
constraint to numerical precision. Maximum deviations are $2.4\times10^{-7}$ and
$4.77\times10^{-7}$ in the three-dimensional eight- and sixteen-grain cases. The hard state
update enforces this property after every neural step (\smsec{i}).

\subsection{Scalar interface motion and topology change}
\label{sec:results:scalar}

Scalar rollouts reproduce curvature-driven shrinkage, sequential extinction, and coalescence
without explicit grain identities (Figure~\ref{fig:scalarladder}). A single grain follows the
radius law of Equation~\eqref{eq:radiuslaw}; two separated grains disappear in the order
$2\rightarrow1\rightarrow0$, with full-field MAE $7.35\times10^{-4}$. Interacting grains
coalesce before the merged domain shrinks; the coalescence time differs by seven simulation-time
units, while the merged-domain extinction error is $0.197\%$. Full errors and event times are in
Table~\ref{tab:headline} and \smsec{b}.

The sparse multigrain case closes within its prescribed tolerance: its residual radius is
$0.014663$ against a $0.025$ limit. Denser scalar fields lose connected components as they
coarsen, but these component counts are not persistent grain identities (Figure~\ref{fig:scalarladder}).

\skelfig{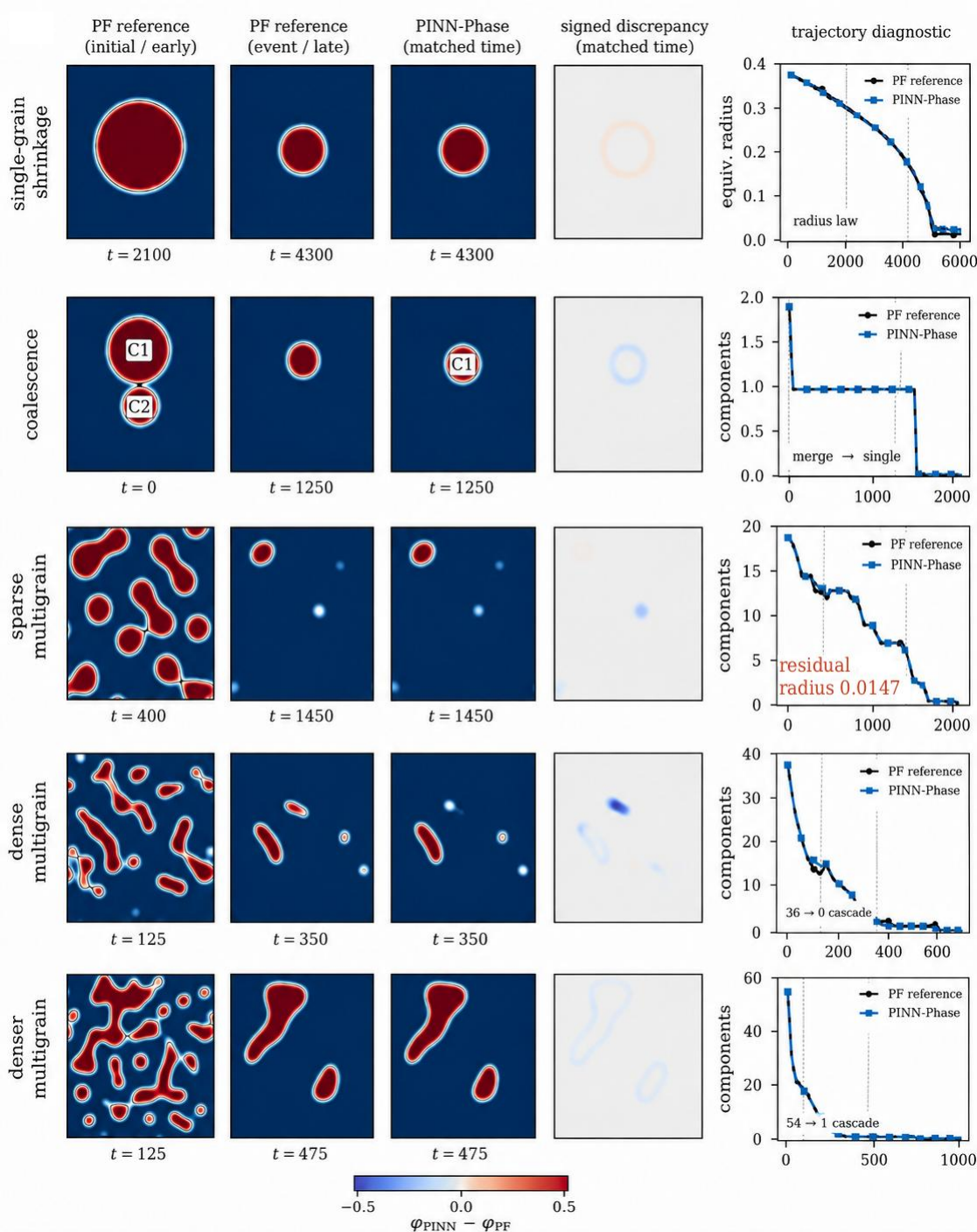}{%
\textbf{Scalar benchmark progression from shrinkage to multicomponent coarsening.} Rows show
single-grain shrinkage, two-grain coalescence, sparse multigrain closure, and two denser scalar
fields. Columns pair the direct phase-field reference with PINN-Phase at the matched time, followed
by signed field error and either equivalent radius or thresholded connected-component count.
The component count describes topology within a scalar field, not persistent grain identity.
The sparse case meets its residual-radius tolerance; full errors and event times appear in
Table~\ref{tab:headline} and \smsec{b}.}{fig:scalarladder}

The predicted diffuse-interface profile and morphology diagnostics follow the reference
(Supplementary Figure~\ref{fig:materials}). Component counts and interface and energy diagnostics
decrease as the area distribution shifts toward fewer, larger domains. These finite simulations do
not establish an asymptotic grain-growth distribution. The sparse-to-dense warm start has elevated
interface-weighted error, and its denser transfer is neutral (Supplementary Table~\ref{tab:s_transfer}).

\subsection{Multiphase relaxation and two-dimensional coarsening}
\label{sec:results:voronoi}

The initially right-angled four-phase junction relaxes toward the equal-energy geometry, with an
RMS angle deviation of $5.5^{\circ}$ at step 6000 (Supplementary Figure~\ref{fig:triplejunction}).
The longest-horizon model retains all four phases and reaches $3.272\%$ label disagreement,
versus $6.543\%$ for the unchanged initial state. Its energy diagnostic is within $0.4\%$ of
the reference; junction angles and full diagnostics are in \smsec{c}.

\mrrev{The two longer-horizon stages show lower late-time disagreement than the shorter stages,
although the intermediate stages are not monotonic.} Epoch count also changes across the
warm-started sequence, so this association cannot isolate a horizon effect (\smsec{c}).

\begin{nthreerev}
The 25-grain cascade retains the correct final survivors through repeated topology changes
\crev{(Figure~\ref{fig:voronoi25})}. The same permutation-equivariant model evaluates a training
initial condition and one held out from training but inspected during development; neither is
prospective evidence. After 12,000 autonomous steps, disagreement is $1.46\%$ and $1.13\%$,
versus $44.37\%$ and $51.25\%$ for the unchanged initial states. Both final survivor sets are
exact, including all thirteen extinctions in the held-out sequence. The last 7,904 steps extend
beyond the represented training horizon; intermediate scores and structural checks are in
Table~\ref{tab:headline} and \smsec{c}.

Event timing remains weaker than final boundary placement. Only three of seven training-case
extinctions inside the represented horizon meet the $\pm50$-step tolerance; offsets are
predominantly premature in both development cases (\smsec{c}).

The hybrid variant instead retains 17 phases where its original 25-grain reference retains 14,
despite improving label agreement over the unchanged initial state. Its main residual is an
extinction-count lag, whereas event timing dominates the cascade-family error. Model variant and
initial condition both differ, so this contrast cannot identify the cause (Supplementary
Figures~\ref{fig:gnn} and \ref{fig:smc_hybrid}).
\end{nthreerev}

\begin{figure}[!tbp]
  \centering
  \IfFileExists{Figure_4.pdf}{%
    \includegraphics[width=\linewidth,height=0.71\textheight,keepaspectratio]{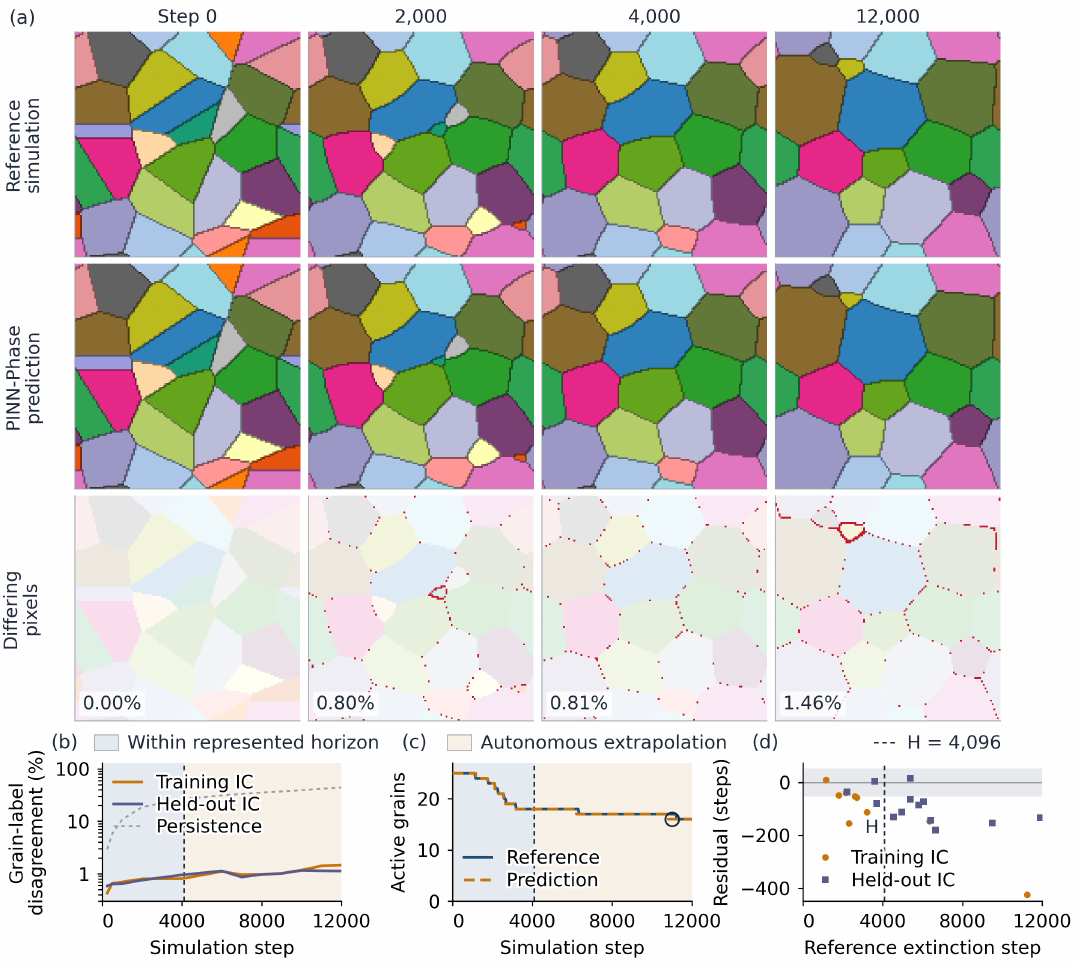}%
  }{\fbox{\parbox[c][0.22\textheight][c]{0.88\linewidth}{\centering\small Pending rendered figure:\par\nolinkurl{Figure_4.pdf}}}}
  \caption{%
\footnotesize
\frev{\textbf{Long-horizon coarsening of a 25-grain cascade.} Both initial conditions are
development cases evaluated with one unchanged permutation-equivariant model. (a) Direct
phase-field reference, PINN-Phase, and differing grain-label pixels at steps 0, 2,000, 4,000,
and 12,000 for the training initial condition. (b) Label disagreement for training and held-out
development conditions, compared with their unchanged initial states. Shading separates the
represented horizon $H=4{,}096$ from autonomous continuation. (c) Active-grain counts for the
trajectory in (a). (d) Extinction-time offsets for both conditions; the dashed line marks $H$
and shading marks the $\pm50$-step tolerance. Offsets are predominantly premature.}}
  \label{fig:voronoi25}
\end{figure}

\begin{genrev}
\mrrev{The dense 64-grain benchmark extends the two-dimensional evaluation to a larger phase
count and repeated extinctions.} The graph-free permutation-equivariant model was scored once at
its 4,096-step represented horizon. A subsequent longer-horizon continuation is a retrospective
sensitivity study, not an independent confirmation (Table~\ref{tab:s_n64_horizon}; Supplementary
Figure~\ref{fig:voronoi64}).
\end{genrev}
\begin{e1rev}
The retrospective continuation recovers all 21 reference survivors and all 43 extinction
identities at step 12,000, with $3.97\%$ label disagreement. The primary evaluation instead
retains one extra grain and ends at $6.09\%$ disagreement. Both outcomes and intermediate scores
are in Table~\ref{tab:s_n64_horizon}. Because the continuation changes both horizon and training
duration after the primary score was known, its improvement cannot be attributed to horizon alone.
\end{e1rev}

\begin{genrev}

\subsection{\mrrev{Prospective transfer to unseen two-dimensional initial conditions}}
\label{sec:results:generalisation}

The preceding development cases motivate a prospective test of transfer within the 25-grain
cascade family. The evaluation uses eight unseen initial conditions and two stress cases fixed
before scoring (Section~\ref{sec:methods:generalisation}).

The model meets all predefined criteria in seven of eight unseen cases and both stress cases
(Figure~\ref{fig:n25gen}; Table~\ref{tab:n25gen}). Final label disagreement spans $0.97\%$ to
$3.71\%$ across the unseen cases, compared with $32.50\%$ to $49.33\%$ for their unchanged
initial states. All ten cases remain structurally valid, and differing pixels lie within the
diffuse-interface region. Supplementary Tables~\ref{tab:s_n25gen_criteria} and
\ref{tab:s_n25gen_cases} give every score; Supplementary
Figures~\ref{fig:sm_gen_atlas}--\ref{fig:sm_gen_atlas_e} show all trajectories.

\mrrev{The seven-of-eight result meets the fixed cohort threshold exactly, without margin.}
The 95\% Wilson interval is $0.529$ to $0.978$, so this small cohort does not resolve the
underlying pass rate precisely. The failed case misses only the survivor-identity criterion: a
grain absent from the prediction remains in the reference at step 12,000. The reference loses it
933 steps later, which does not change the scored failure (\smsec{g}). A prespecified descriptive
second model passes and fails the same cases (Supplementary Table~\ref{tab:s_n25gen_arms}).

\mrrev{The result supports prospective transfer within this fixed 25-grain family, phase count,
and resolution, without case-specific retraining or tuning.}
\end{genrev}

\begin{table}[!tbp]
\centering
\footnotesize
\caption{\gencell{\mrrev{Prospective transfer to unseen initial conditions of the 25-grain cascade family.} A case counts as reproduced only when it satisfies every criterion fixed in advance (Supplementary Table~\ref{tab:s_n25gen_criteria}). Terminal quantities are at simulation step 12,000.}}
\label{tab:n25gen}
\begin{tabular}{lccc}
\toprule
 & \gencell{Unseen initial conditions} & \gencell{Stress cases} & \gencell{Criterion} \\
\midrule
\gencell{Cases evaluated} & \gencell{8} & \gencell{2} & \gencell{n/a} \\
\gencell{Cases meeting all predefined criteria} & \gencell{7} & \gencell{2} & \gencell{$\geq 7$ of 8; 2 of 2} \\
\gencell{Wilson 95\% interval on the rate} & \gencell{0.529--0.978} & \gencell{n/a} & \gencell{n/a} \\
\gencell{Grain-label disagreement (\%)} & \gencell{0.97--3.71} & \gencell{0.94--2.15} & \gencell{$\leq 15$} \\
\gencell{Persistence baseline (\%)} & \gencell{32.50--49.33} & \gencell{35.49--48.86} & \gencell{$\geq 30$} \\
\gencell{Gain over persistence, step 12{,}000} & \gencell{0.912--0.975} & \gencell{0.956--0.974} & \gencell{$\geq 0.60$} \\
\gencell{Terminal grain-identity $F_1$} & \gencell{0.968--1.000} & \gencell{1.000--1.000} & \gencell{$\geq 0.90$} \\
\gencell{Cases missing a reference survivor} & \gencell{1} & \gencell{0} & \gencell{0 required} \\
\gencell{Cases with a grain reappearing} & \gencell{0} & \gencell{0} & \gencell{0 required} \\
\bottomrule
\end{tabular}
\par\vspace{2pt}\footnotesize\raggedright
\gencell{Per-case values are in Supplementary Table~\ref{tab:s_n25gen_cases}. The ten cases and all thresholds were fixed before scoring, and scoring was performed once. Extinction timing was declared descriptive in advance and enters no criterion. This cohort is never pooled with the earlier partial cohort of Section~\ref{sm:g}.}
\end{table}

\skelfig[0.94\linewidth]{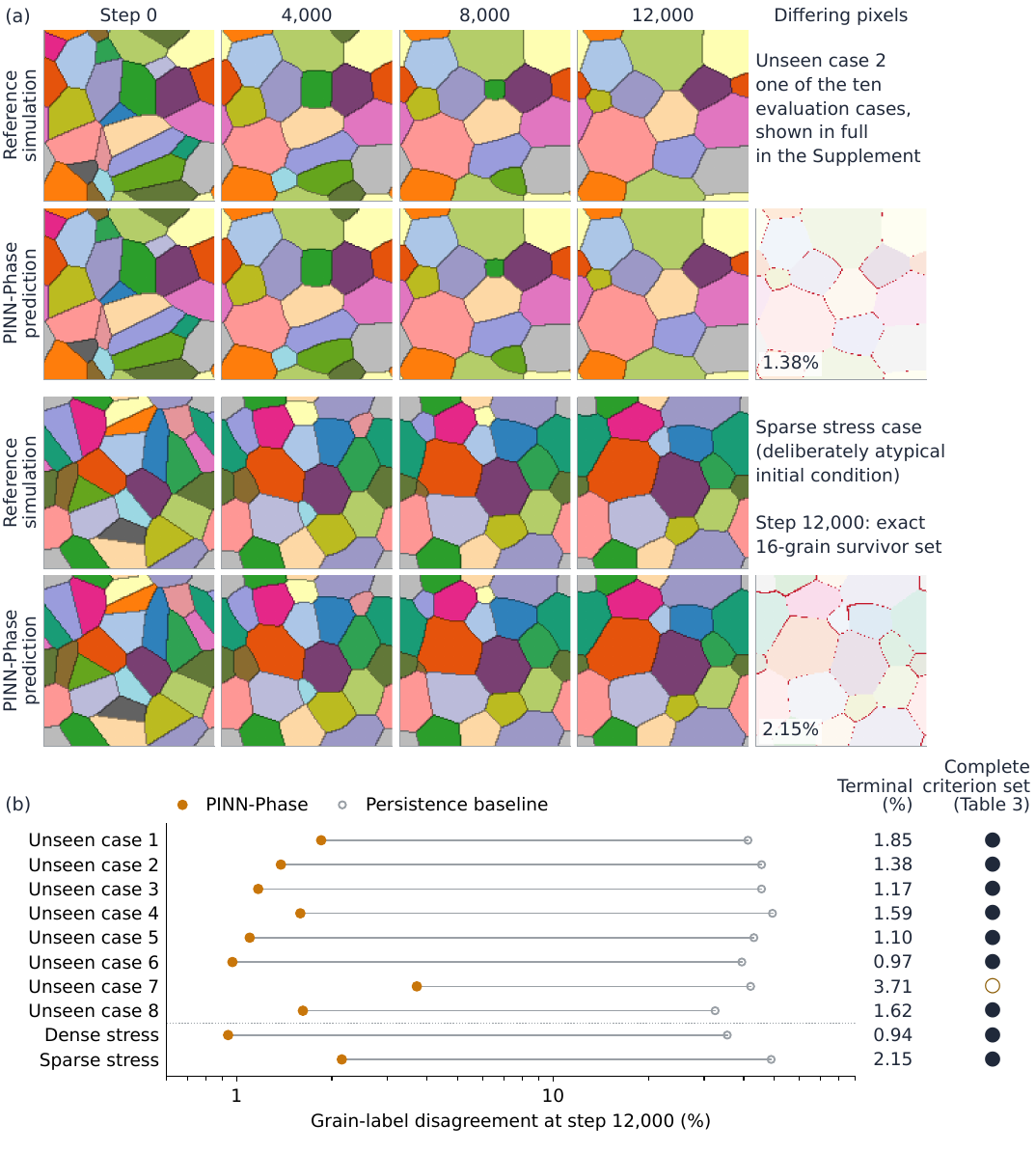}{%
\gencell{\textbf{Prospective transfer to unseen two-dimensional grain structures.}
(a) Direct phase-field reference and unchanged-weight PINN-Phase for unseen case 2 and the sparse
stress case at steps 0, 4{,}000, 8{,}000, and 12{,}000. The last column marks differing grain
labels; colors denote identity within each case. (b) Final label disagreement for all eight
unseen cases and both stress cases in fixed evaluation order, compared with the unchanged-initial-state
baseline on a logarithmic axis. Filled markers denote PINN-Phase and open markers the baseline;
the right column reports complete-criterion status. Seven unseen cases and both stress cases
meet all criteria in Table~\ref{tab:n25gen}. Full histories appear in Supplementary
Figures~\ref{fig:sm_gen_atlas}--\ref{fig:sm_gen_atlas_e}.}}{fig:n25gen}

\subsection{Three-dimensional progression}
\label{sec:results:3d}

Three-dimensional evolution is first established for scalar spherical shrinkage, where explicit
bound enforcement is required (Supplementary Figure~\ref{fig:3dscalar}). In the reported bounded
configuration, the equivalent-radius slope is $0.991$ times the reference slope and the terminal
residual radius is $0.034515$; a clamp-only diagnostic gives $0.011613$, and the unbounded control
becomes inadmissible. The outcome therefore depends on the bound-enforcement scheme.

In the first explicit MPF cube ($64^3$, eight grains, one designed extinction at step 1800),
PINN-Phase trained from the initial cube with a horizon curriculum to 2048 steps reaches $99.53\%$
label agreement at step 2000, against $94.01\%$ for persistence. It recovers the exact seven-grain
survivor set and removes the designed grain one saved frame after the reference; at step 3200,
agreement remains $99.15\%$ against $92.82\%$ (Supplementary Figures~\ref{fig:n8cube_ref}
and~\ref{fig:n8cube_learned}). This is a single designed-extinction benchmark, not a statistical
test of grain-growth kinetics.

The second cube ($96^3$, 16 grains) contains three designed interior extinctions, of grains 9, 10
and 11 at steps 1000, 1200 and 1400 (Supplementary Figure~\ref{fig:n16cube_ref}). \mrrev{The final
model continues an earlier $H=2048$ checkpoint for 20 epochs with the strict clip-and-renormalize
map in training and prediction.} It reproduces the three extinctions at the same saved frames as the
reference and recovers the exact thirteen-grain survivor set (Figure~\ref{fig:n16cube_learned}).
Terminal label agreement is $99.668\%$ against $93.32\%$ for persistence; the remaining $0.3315\%$
disagreement lies within one voxel of the reference interface, and the unit-sum error stays at most
$4.77\times10^{-7}$ through step 3200. Event times are exact at the 200-step saved-frame resolution.
A long-horizon check found that the dense $N=16$ reference formulation loses phase purity at later
times, so these results are not evidence of sustained sharp-field coarsening (Supplementary
Section~\ref{sm:i}).

\skelfig{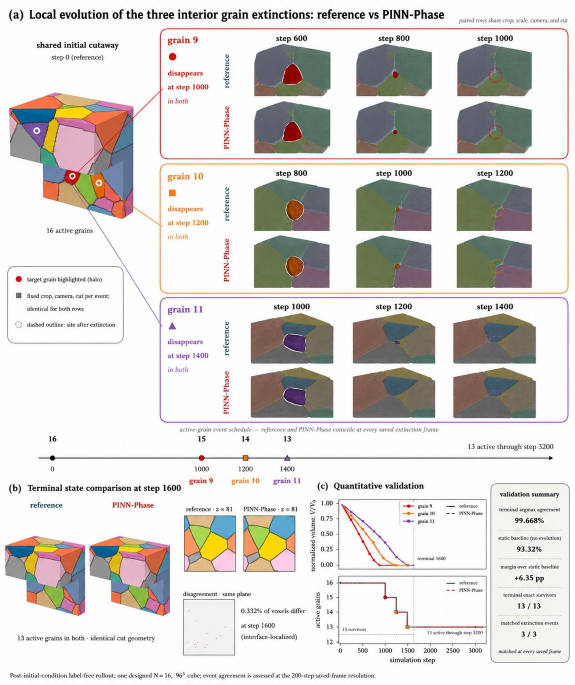}{%
\textbf{Prediction of three interior grain extinctions in the $N=16$, $96^3$ cube.}
(a) Reference and PINN-Phase neighborhoods for grains 9, 10, and 11 at matched saved frames.
(b) Terminal cutaways at step 1600 and the interface-localized label mismatch. (c) Normalized
target-grain volumes and active-grain count. The model reproduces the $16\rightarrow15\rightarrow14\rightarrow13$
sequence at the 200-step saved-frame resolution, recovers the exact terminal survivor set, and
reaches $99.668\%$ label agreement, $6.35$ percentage points above persistence.}{fig:n16cube_learned}

\begin{n16rev}
\subsection{Prospective transfer to unseen three-dimensional initial conditions}
\label{sec:results:n16cohort}

With one unchanged model, five of six unseen microstructures of the same designed-extinction family
meet the complete predefined qualification (Figure~\ref{fig:n16cohort}; Table~\ref{tab:n16cohort}).
The six cases were fixed before evaluation and advanced without case-specific retraining, tuning or
checkpoint selection. At step 1600, label agreement is $95.06\%$ to $95.83\%$, $+1.71$ to $+2.47$
percentage points above each case's persistence baseline, and all six recover the exact
thirteen-grain active set and the identities of all three disappearing grains. All eighteen
extinction events lie within one 200-step output interval of the reference, and eleven coincide
exactly.

The remaining case, unseen microstructure 5, passes terminal topology, extinction identity and
timing, but its gain over persistence becomes slightly negative at the final saved state
($-0.081$ percentage points at step 3200), while the nearest qualifying case stays positive by
$+0.007$ points. \mrrev{The two cases lie on opposite sides of a numerically marginal criterion}
(Supplementary Figure~\ref{fig:n16cohortsm}). The result supports prospective transfer within this
\mrrev{designed-extinction family}, with the phase-purity caveat of Section~\ref{sec:results:3d}.

\skelfig{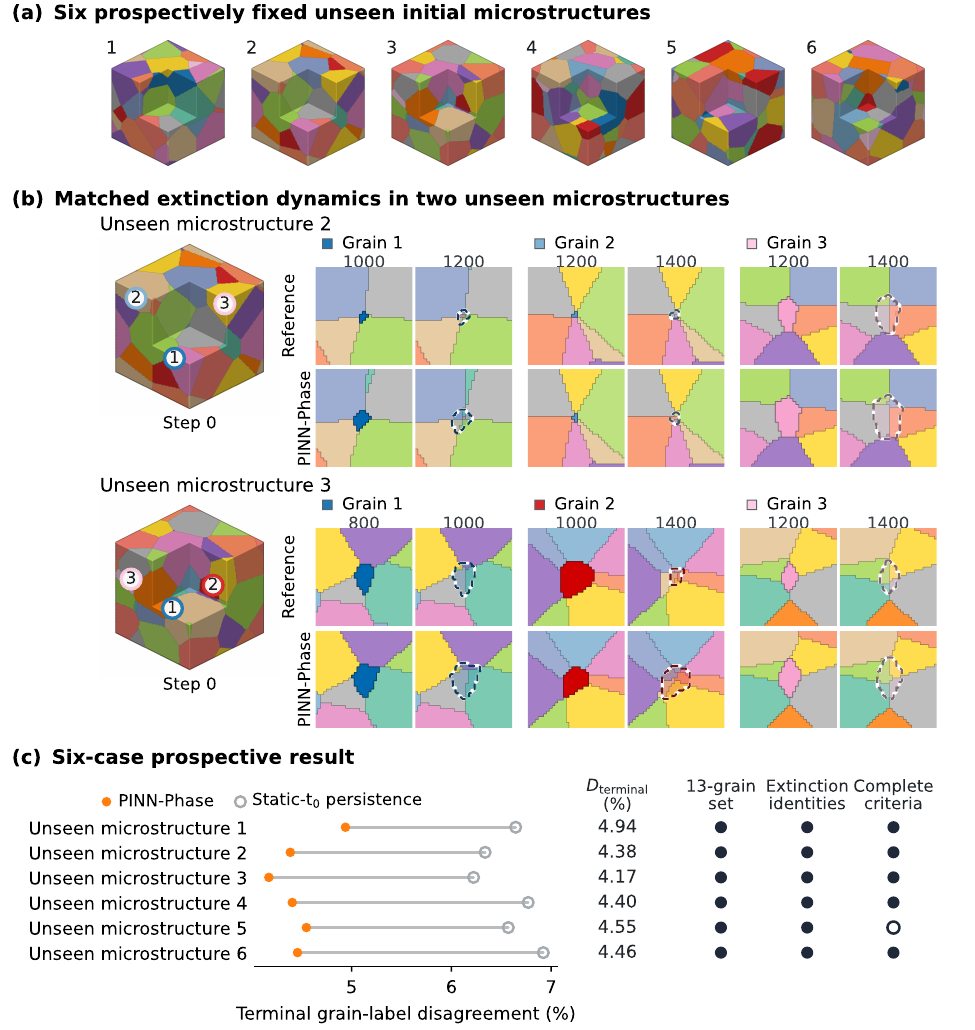}{%
\nsixteencell{\textbf{Prospective transfer to unseen three-dimensional initial conditions.}
(a) Six unseen $N=16$, $96^3$ microstructures of the same designed-extinction family, advanced
with one unchanged model. Colors denote grain identity. (b) Extinction dynamics in two successful
illustrative cases. Grain 1--3 gives reference disappearance order; dashed contours mark a grain's
last occupied region after extinction. One event in the second case differs by one saved interval.
(c) Terminal grain-label disagreement $D_\mathrm{terminal}$, the voxel fraction with a different
argmax identity from the reference at step 1600, compared with each case's static-$t_0$
persistence baseline. Symbols also show terminal active-set identity, extinction-identity recovery
and complete-criterion status. Monitoring continues to step 3200. All 6 cases recover the exact
thirteen-grain terminal set and three extinction identities; 5 meet the complete qualification.}}{fig:n16cohort}
\end{n16rev}

\subsection{Native \texorpdfstring{$128^3$}{128\textthreesuperior} long-horizon evolution after multi-initial-condition training}
\label{sec:results:n128}

On its training microstructure, the multi-initial-condition model stays close to the reference far
beyond its training horizon (Figure~\ref{fig:n128cube}). Advanced from $H=4{,}096$ to $T=24{,}000$
steps without reference corrections, it ends with $1.2122\%$ label disagreement ($98.79\%$ agreement,
against $69.48\%$ for persistence). It retains the exact five-grain survivor set and the reference
extinction sequence, with event residuals of $-20$, $-100$ and $+40$ steps. The prediction is
closer to the reference than persistence at every saved state. This is temporal extrapolation on a
microstructure seen in training, not initial-condition transfer. The earlier single-initial-condition
model reached $0.9786\%$ terminal disagreement on this field (Supplementary
Figure~\ref{fig:historical128}). The two training histories differ, so this comparison does not
isolate an effect of adding training geometries (Supplementary Section~\ref{sm:k}).

\skelfig[0.98\linewidth]{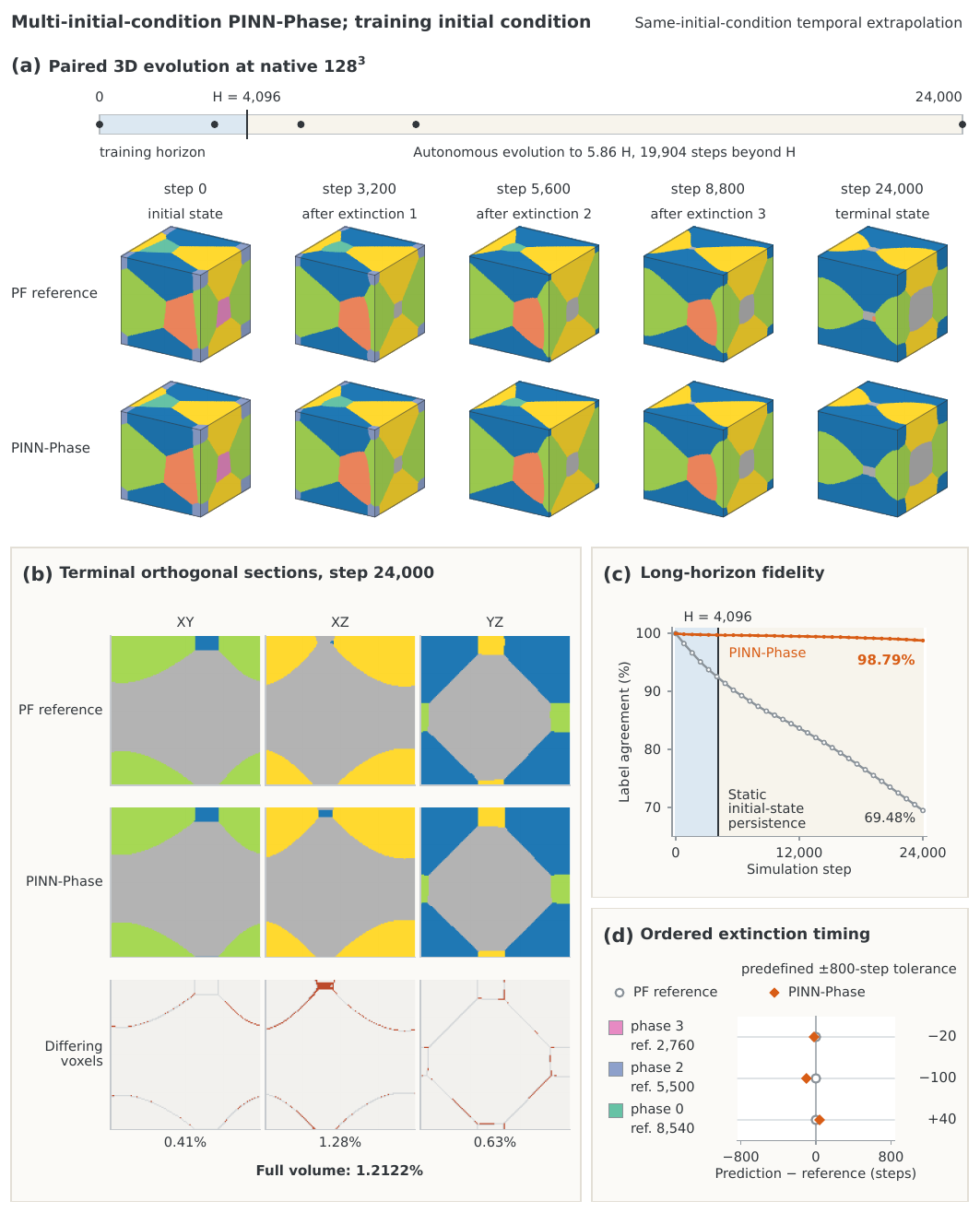}{%
\textbf{Native $128^3$ evolution on the training initial condition.}
(a) Phase-field reference and multi-initial-condition PINN-Phase predictions through
$T=24{,}000=5.86H$, where $H=4{,}096$ is the represented training horizon.
(b) Terminal orthogonal sections and differing pixels (percentages below the sections are
slice-specific; full-volume disagreement $1.2122\%$).
(c) Label agreement against persistence.
(d) Signed extinction-time differences at 20-step ledger resolution.
This comparison tests same-initial-condition temporal extrapolation.}{fig:n128cube}

\subsection{Initial-condition transfer at native \texorpdfstring{$128^3$}{128\textthreesuperior}}
\label{sec:results:n128transfer}

The same model transfers to unseen $128^3$ microstructures without case-specific optimization. Two
cohorts of six initial conditions are reported separately: a development cohort inspected during
development, and a prospective unseen cohort fixed before any of its outcomes were opened
(Figure~\ref{fig:n128transfer}; Supplementary Sections~\ref{sm:l} and~\ref{sm:n}). Within the training horizon, at step 4,000, the prospective cases reach $0.72$--$2.11\%$ label
disagreement, against $6.69$--$7.90\%$ for persistence. Their first extinction differs from the
reference by at most 180 steps. The development cohort gives $0.84$--$1.00\%$ against
$6.66$--$8.25\%$.

At step 24,000, nearly six times the training horizon, all six prospective predictions remain closer
to the reference than persistence at every saved state, and five recover the exact terminal
survivor set with $2.84$--$4.72\%$ disagreement. The remaining case, unseen microstructure 4,
reaches $11.30\%$ and retains one grain that has disappeared in the reference. The other long-horizon
discrepancies concern extinction timing, both early and late rather than a common clock offset
(Supplementary Section~\ref{sm:l}). All development cases retain the reference terminal survivor set,
with exceptions confined to final-event timing, whereas the single-initial-condition predecessor had
missed the field and timing criteria on all six. Strict event timing is therefore the most demanding
quantity when transfer is combined with long temporal extrapolation.

\skelfig[0.98\linewidth]{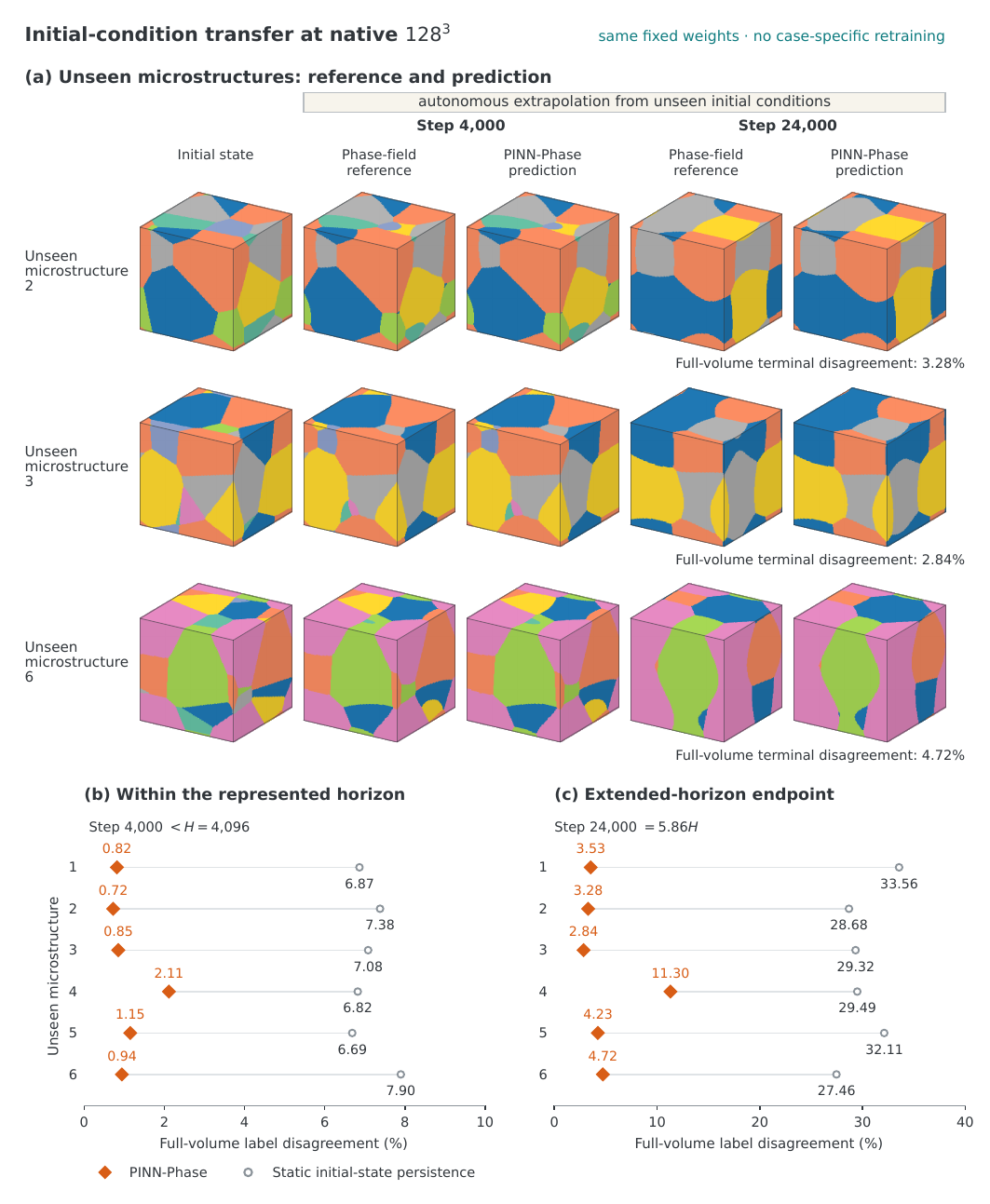}{%
\textbf{Initial-condition transfer at native $128^3$.}
(a) Prospective unseen microstructures 2, 3 and 6: each initial state and paired phase-field
reference and PINN-Phase predictions at step 4,000 and at the 24,000-step extended-horizon
endpoint.
(b,c) Full-volume label disagreement for all six prospective unseen microstructures against
static-initial-state persistence at step 4,000, within the represented training horizon
$H=4{,}096$, and at $T=24{,}000=5.86H$.
Panel (a) shows three illustrative cases; panels (b,c) report all six. Microstructure 4
retains one additional terminal survivor (Supplementary Section~\ref{sm:l}).}{fig:n128transfer}

\subsection{Case-specific physics-informed continuation}
\label{sec:results:specialization}

Case-specific physics-informed continuation improved extinction timing in two previously inspected
development microstructures (Supplementary Figure~\ref{fig:specialization}). Starting from the
trained weights, each model received 1,024 additional updates using only its initial condition and
the governing physical residual. Both previously failing cases then met the predefined timing
tolerance, with lower final label disagreement; in development microstructure 5, for example, the
final extinction error decreased from $+1{,}460$ to $+520$ steps. Development microstructure 4,
which already passed, retained all six criteria. Individual grain-volume histories did not improve
uniformly. Because the cases were selected after their earlier outcomes were known, these
retrospective tests demonstrate a possible repair strategy, not a prospective adaptation policy;
the full before-and-after values are in Supplementary Section~\ref{sm:m}.

\begin{figure}[p]
\centering
\includegraphics[width=\linewidth,height=0.83\textheight,keepaspectratio]{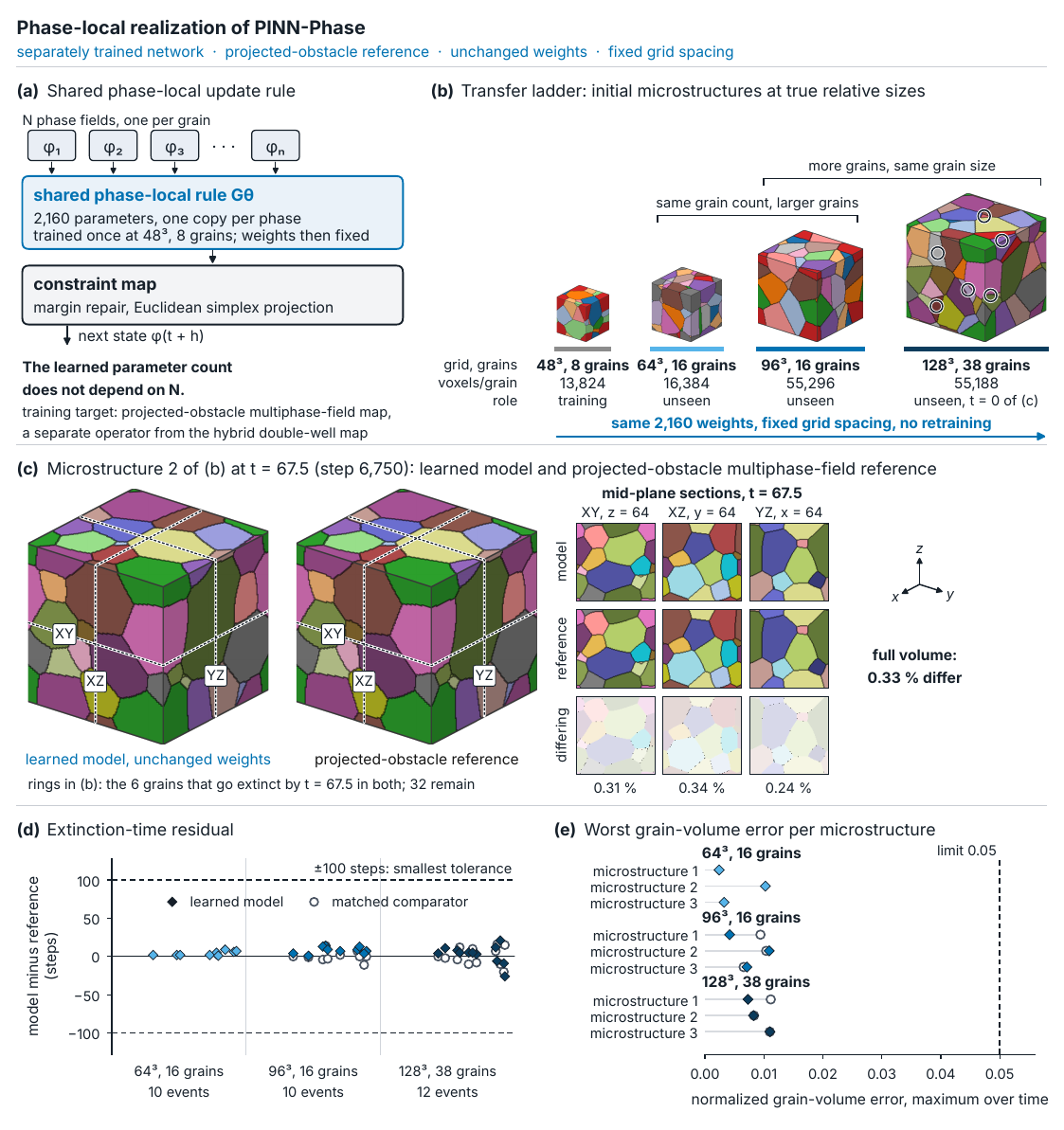}
\caption{\footnotesize\textbf{Phase-local transfer across domain size and grain count with unchanged weights.}
(a)~One shared rule $G_\theta$ (2,160 parameters, independent of the number of grains $N$) acts on each
phase field; margin repair and a joint Euclidean simplex projection then couple all phases at every
voxel. The rule was trained once at $48^3$ with eight grains against one-step targets of the
projected-obstacle reference operator. (b)~Training microstructure and one unseen initial
microstructure per regime at true relative size, with the mean initial grain volume below each; each
regime has three unseen microstructures. Rings mark the six grains of the $128^3$ case that disappear
within the horizon. (c)~That case (microstructure 2) at $t=67.5$: model and reference argmax-label
cubes and center-plane sections, with differing voxels in black ($0.33\%$ of the volume). (d)~Signed
extinction-time residuals, model minus reference step (10, 10 and 12 events): learned model filled,
matched comparator open (not evaluated at $64^3$); dashed lines at 100 steps, the lower bound of the
tolerance $\max(100,\,0.1\,s_{\mathrm{ref}})$. (e)~Maximum normalized grain-volume error of each
microstructure; limit $0.05$. Extinctions and volumes use the diffuse fields. All $128^3$ cases are
shown in Supplementary Figures~\ref{fig:sm_phaselocal_cases} to~\ref{fig:sm_phaselocal_parity}.}
\label{fig:phaselocal}
\end{figure}

\subsection{Transfer across domain size and grain number}
\label{sec:results:phaselocal}

With unchanged weights, the phase-local realization transfers to larger three-dimensional domains
with more grains (Figure~\ref{fig:phaselocal}). The 2,160-parameter checkpoint, trained once at
$48^3$ with eight grains, meets all acceptance criteria on the three unseen microstructures of each
regime, $64^3$ with 16 grains, $96^3$ with 16 grains and $128^3$ with 38 grains, at fixed grid spacing.
All 32 reference extinctions are reproduced by identity, with residuals of at most 26 steps against a
tolerance of at least 100 steps (Figure~\ref{fig:phaselocal}d). Alive sets agree at every evaluation
time, no extinct phase returns, and the maximum normalized grain-volume error is at most $0.0110$
against the limit $0.05$ (Figure~\ref{fig:phaselocal}e). The continuous-field error stays at most
$0.0114$ times that of static persistence. The regimes first increase the mean grain volume at fixed
phase count ($13{,}824$ voxels per grain at training, $55{,}296$ at $96^3$). They then increase the
number of grains at nearly fixed mean volume, without isolating either factor. The largest system
holds about $90$ times as many phase-field values as a training field. Per-regime values are in
Supplementary Table~\ref{tab:s_headline_full} and Section~\ref{sm:phaselocal}.

A matched comparator with the same architecture, loss and update budget, but trained on
reference-trajectory states, also meets all casewise criteria at $96^3$ and $128^3$ (it was not
evaluated at $64^3$; Supplementary Section~\ref{sm:phaselocal}). Training on the model's own states
is therefore viable, not shown to be superior, and the transfer cannot be attributed to that choice.

In the median $128^3$ case (Figure~\ref{fig:phaselocal}c), model and reference agree in grain
identities and morphology at $t=67.5$: $0.334\%$ of the voxels differ in label, and the six
extinctions occur $+4$ to $+11$ steps after the reference. All three $128^3$ cases are shown in
Supplementary Figures~\ref{fig:sm_phaselocal_cases} to~\ref{fig:sm_phaselocal_parity}. The cases
were admitted because their reference evolution contains extinctions, so these outcomes do not
estimate performance on unselected microstructures; within this scope
(Section~\ref{sec:discussion:scope}), the three regimes establish unchanged-weight transfer across
the tested domain sizes and grain counts in one synthetic geometry family.

\begin{figure}[!tbp]
\centering
\IfFileExists{Figure_11.pdf}{%
  \includegraphics[width=\linewidth,height=0.50\textheight,keepaspectratio]{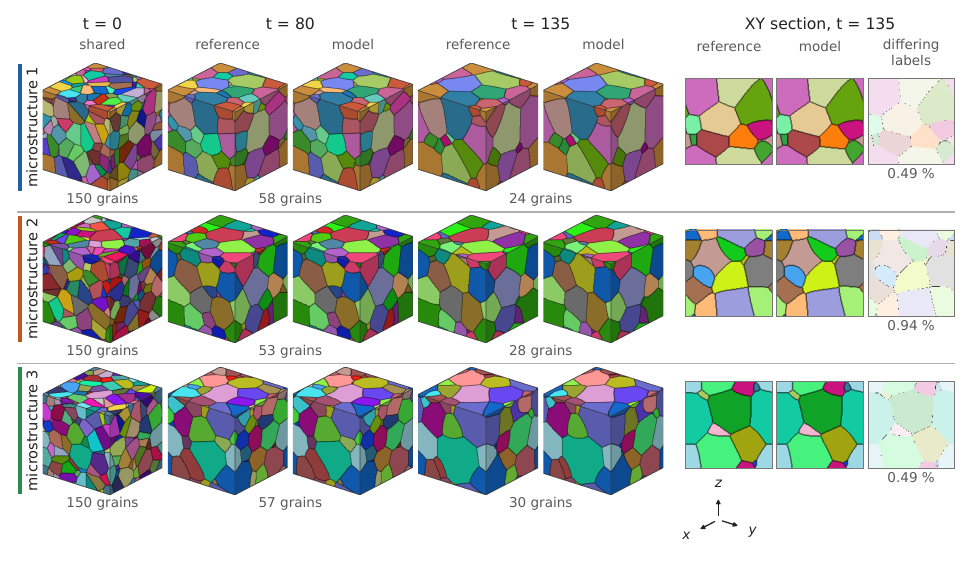}%
}{\fbox{\parbox[c][0.44\textheight][c]{0.95\linewidth}{\centering\small figure pending\par\nolinkurl{Figure_11.pdf}}}}
\caption{\footnotesize\textbf{Transfer to $128^3$ with 150 grains in three microstructures.}
The unchanged checkpoint of Figure~\ref{fig:phaselocal} advances each microstructure (rows 1 to 3)
for 13,500 steps from a shared initial state. Reference and model argmax-label cubes are shown at
$t=80$ and $t=135$, and XY mid-plane sections ($z=64$) at $t=135$ mark differing labels in black on
the faded reference section. Printed counts are alive grains (diffuse volume above $10^{-6}$;
Section~\ref{sec:methods:phaselocal}), equal in reference and model; all 126, 122 and 120 reference
extinctions are matched by identity. At $t=135$, $0.51\%$, $0.70\%$ and $0.88\%$ of the whole volume
carry a different label. Colors identify grains; the key gives the cube axes. All four stored times
appear in Supplementary Figure~\ref{fig:sm_n150_cases}.}
\label{fig:n150micro}
\end{figure}

\begin{figure}[!tbp]
\centering
\IfFileExists{Figure_12.pdf}{%
  \includegraphics[width=\linewidth,height=0.48\textheight,keepaspectratio]{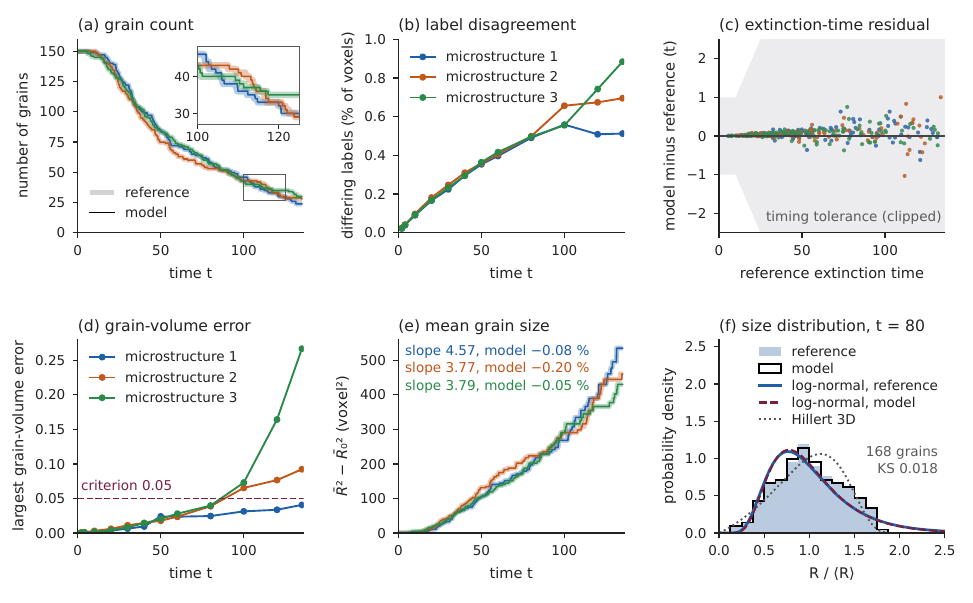}%
}{\fbox{\parbox[c][0.44\textheight][c]{0.95\linewidth}{\centering\small figure pending\par\nolinkurl{Figure_12.pdf}}}}
\caption{\footnotesize\textbf{Event, field and grain statistics for 150 grains.}
(a)~Alive grains; the inset magnifies $100\le t\le125$ (box), the window with the largest mean
reference--model difference in alive count (at most two grains at any time). (b)~Whole-volume label disagreement at twelve evaluation
times. (c)~Model-minus-reference extinction time for 368 events; shaded tolerance
$\max(1,\,0.1\,t_{\mathrm{ref}})$ is clipped at the axis limits. (d)~Largest grain-volume error
relative to mean initial grain volume; dashed, the $0.05$ criterion. (e)~$\bar R^2-\bar R_0^2$; printed are
reference fit slopes over $40\le t\le135$ and model deviations. $\bar R$ uses fixed domain
volume and alive grain count. In (a) and (e), wide translucent lines are reference; thin lines of
the same color are model.
(f)~Radii $R/\langle R\rangle$ at $t=80$, normalized by each
microstructure's mean radius and pooled (168 grains above one voxel): reference, model,
log-normal fits and unit-mean three-dimensional Hillert distribution. KS is the two-sample
Kolmogorov--Smirnov distance.}
\label{fig:n150stats}
\end{figure}

\subsection{Transfer to 150 grains at comparable mean grain volume}
\label{sec:results:n150}

A fourth regime keeps the mean initial grain volume comparable to training ($13{,}981$ against
$13{,}824$ voxels) and enlarges domain and grain count. At $128^3$ with 150 grains, it has $18.96$
times the training domain volume, $18.75$ times its grain count and about 356 times its phase-field
values. Three admitted unseen microstructures are advanced autonomously for 13,500 steps ($t=135$),
4.5 times the training rollout length (Figures~\ref{fig:n150micro} and~\ref{fig:n150stats}). The
reference eliminates 126, 122 and 120 grains, leaving 24, 28 and 30 at $t=135$. The model reproduces
all 368 extinctions by identity, with no spurious extinction or return, and its grain count follows
the reference throughout (Figure~\ref{fig:n150stats}a). Extinction-time residuals have absolute
medians of 6, 5 and 7 steps and maxima of 63, 103 and 75 steps, all within the tolerance
$\max(100,\,0.1\,s_{\mathrm{ref}})$ (Figure~\ref{fig:n150stats}c). Label disagreement stays at or
below $0.42\%$ up to $t=60$ and is $0.51\%$, $0.70\%$ and $0.88\%$ at $t=135$
(Figure~\ref{fig:n150stats}b). The continuous-field error is at most $0.011$ of persistence at every
evaluation time, against limits of $0.5$ and, at $t=135$, $0.25$. All states are admissible
(unit-sum error at most $10^{-12}$), and the structural check passes.

Grain identities, extinction timing within tolerance and field agreement are thus reproduced
throughout. The per-grain volume criterion holds up to $t=80$ in all three microstructures, beyond the
6,750-step horizon of the $128^3$/38-grain regime, and up to $t=135$ in microstructure 1. The largest
normalized individual grain-volume error stays at or below $0.040$ up to $t=80$. It reaches $0.065$
and $0.073$ at $t=100$ in microstructures 2 and 3, and is $0.041$, $0.092$ and $0.27$ at $t=135$
(Figure~\ref{fig:n150stats}d). The alive set equals the reference at all twelve times in
microstructure 2. In microstructures 1 and 3, it differs by one grain at one instant, $t=50$ and
$t=40$, respectively. The model and reference extinction steps of that grain, 17 and 5 steps apart,
fall on either side of the evaluation time. Both match in identity and within tolerance.

The equivalent radius $\bar R=(3V/(4\pi N_{\mathrm{alive}}))^{1/3}$ uses the fixed domain volume $V$
and the number $N_{\mathrm{alive}}$ of surviving grains. Its parabolic law therefore follows that
count. Fitted over $40\le t\le135$, the model slopes of $\bar R^2-\bar R_0^2$ lie within $-0.08\%$,
$-0.20\%$ and $-0.05\%$ of the reference slopes (Figure~\ref{fig:n150stats}e), reflecting the matched
extinction history. For the grain-scale check,
radii $R$ are normalized by each microstructure's mean radius $\langle R\rangle$ and pooled. The
two-sample Kolmogorov--Smirnov distance is $0.011$, $0.018$
and $0.024$ at $t=30$, $80$ and $135$ (352, 168 and 82 grains). The corresponding log-normal widths
are $0.468/0.485$, $0.434/0.422$ and $0.527/0.594$ (reference/model; Figure~\ref{fig:n150stats}f;
Supplementary Figure~\ref{fig:sm_n150_sizedist} and Section~\ref{sm:n150:grains}). At comparable
mean grain volume, the unchanged rule carries the grain-resolved extinction history and field
agreement across a nearly nineteenfold increase in grain count and domain volume. The per-grain
volume deviations that grow after $t=80$ define its accuracy range at this population size. These results concern three admitted
microstructures of one synthetic geometry family at fixed grid spacing.

\subsection{Transfer to 1,000 and 5,000 grains}
\label{sec:results:population}

With unchanged weights, the phase-local realization reproduces the coarsening of 3D polycrystals
with 1,000 and 5,000 grains grain by grain (Figure~\ref{fig:population}). In three $240^3$ runs with
1,000 grains ($t\le135$), 2,431 of 2,432 reference extinctions are matched by identity within their
timing tolerance. In one $416^3$ run with 5,000 grains ($t\le50$), 2,109 of 2,111 are matched. The
remaining events straddle the horizon: one reference-only and one model-only extinction at 1,000
grains, and two reference-only extinctions at 5,000 grains. The matched offsets have medians of 6 and
3 steps, and label disagreement stays below 1\%. The surviving fraction and mean grain size coincide
with the reference (Figure~\ref{fig:population}b). These runs extend the 150-grain regime at nearly
constant mean grain volume (Section~\ref{sec:methods:population}), to 33 times more grains and 651
times the voxels of a training field. Per-case values are in Supplementary
Table~\ref{tab:s_population_cases}.

The learned rule is local, and grain remapping keeps 21 to 37 phase channels in use from 150 to 5,000
grains. The deployment therefore avoids the memory growth of one channel per grain: the 5,000-grain
run fits on one GPU with 61.7\,GB peak memory. Over a 34-fold increase in voxels and a 33-fold
increase in grains, the measured time per voxel and step of the learned step changes by 8\%
(Figure~\ref{fig:population}c). At fixed grid spacing, the cost of a run follows its grid size rather
than its number of grains. Run times of training and of every rollout, for the learned model and the
reference, are listed in Supplementary Section~\ref{sm:population}.

\skelfig{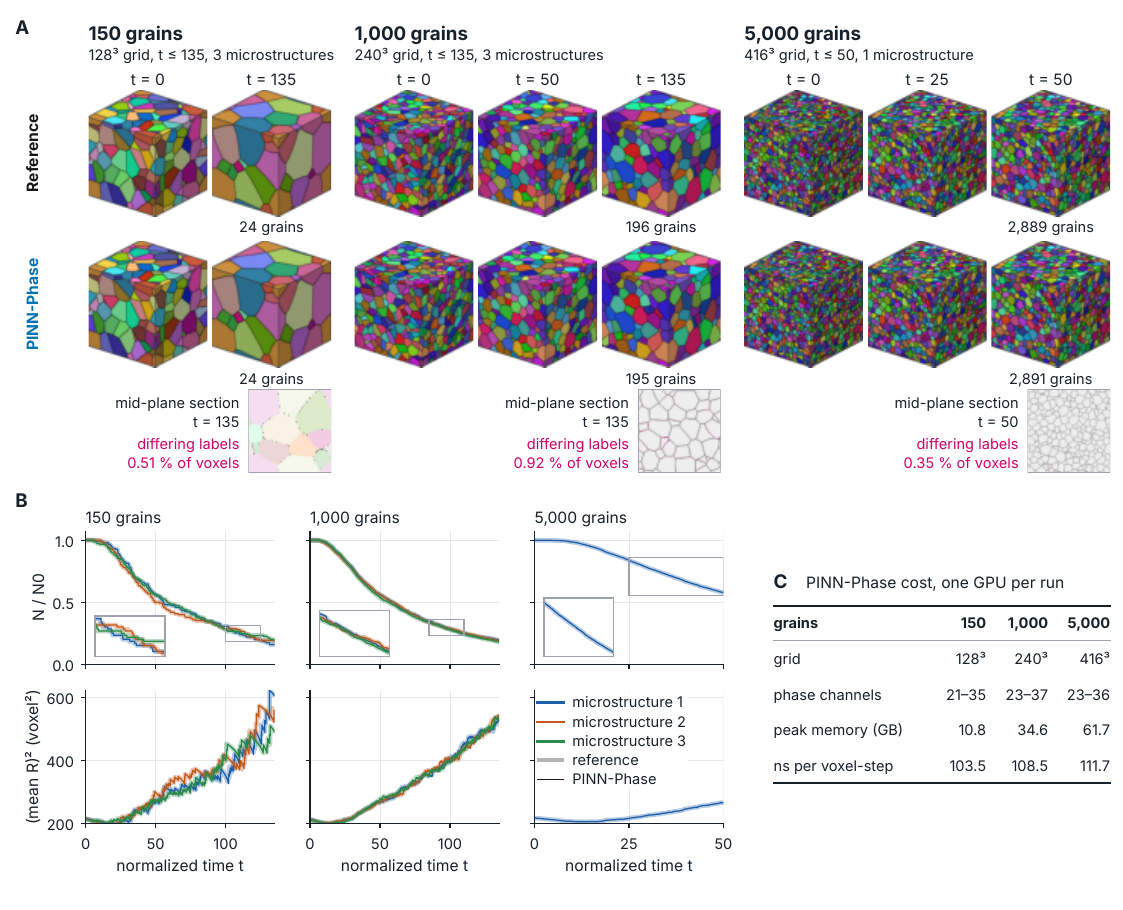}{%
\footnotesize\textbf{Grain growth with unchanged weights at three population sizes.}
(a)~Reference (top) and PINN-Phase (bottom) microstructures with 150, 1,000 and 5,000 grains; the
5,000-grain run is one microstructure to $t=50$. Numbers give the alive grains at the last time
shown; mid-plane sections mark differing labels in magenta, with whole-volume percentages. Colors
denote grain identity, not orientation. (b)~Surviving fraction $N/N_0$ and mean grain size $\bar R^2$
per microstructure (color; reference wide, PINN-Phase thin); insets magnify the boxed windows of
largest mean difference. (c)~Phase channels in use, peak memory and time per
voxel and step of PINN-Phase, one NVIDIA H100 GPU per run; run times of both methods are in
Supplementary Table~\ref{tab:s_population_cost}. Normalized time $t$ is the step number divided by
100.}{fig:population}

\subsection{Grain-resolved fidelity and its limits}
\label{sec:results:fidelity}

Typical per-grain errors change little with the number of grains through the common window
$t\le50$: the 90th percentile is about 1\% of the mean initial grain volume at 150, 1,000 and 5,000
grains (Figure~\ref{fig:fidelity}a, where the exact quantiles are printed). Only the maximum, an
extreme over more grains, grows with the population ($0.024$, $0.032$ and $0.038$) and stays below
the $0.05$ criterion.

The longer 150- and 1,000-grain trajectories define where the criterion ends
(Figure~\ref{fig:fidelity}b). In the three 1,000-grain runs, the maximum error first exceeds $0.05$ at
$t=65$, 60 and 70 and reaches $0.57$, $0.23$ and $0.16$ at $t=135$. At 150 grains it exceeds the
criterion in two of three runs, first at $t=100$ (Section~\ref{sec:results:n150}). The late errors
concentrate in a few large grown grains. At $t=135$, the 77 to 84 grains above five times the mean
initial volume in each 1,000-grain run have median errors of $0.3$ to $0.5\%$ of their own volume;
the worst single grain is off by $10.4\%$ of its volume. Extinction timing does not degrade in the
same way: every matched event lies within its tolerance band at all three sizes
(Figure~\ref{fig:fidelity}c). The strict per-grain volume criterion is therefore met through
$t=57.5$ to $67.5$ at 1,000 grains, and through the full horizon $t=50$ in the one 5,000-grain
microstructure. Beyond that point, grain identities, extinction timing and the typical grain-volume
error remain accurate. The volume of the largest grains defines the accuracy range of the unchanged
weights at these population sizes.

\skelfig{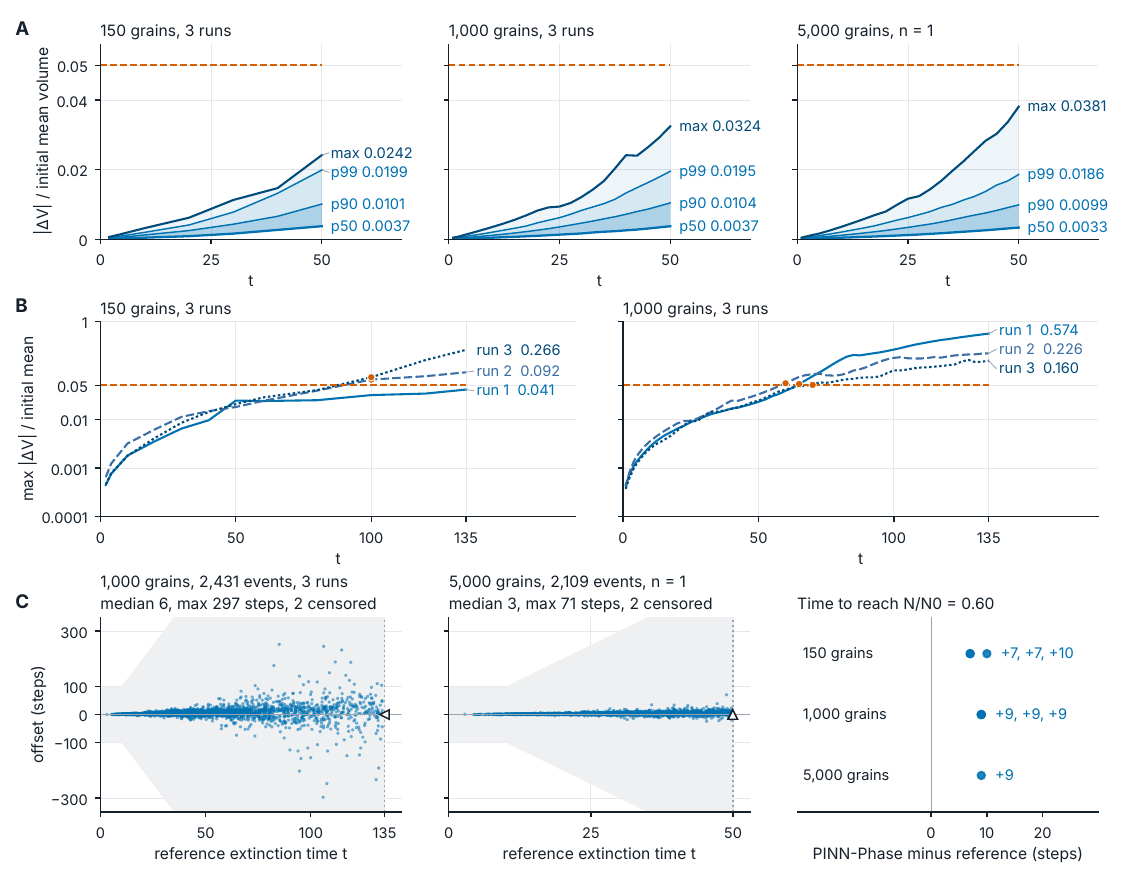}{%
\footnotesize\textbf{Per-grain accuracy and its time limit.}
(a)~Median, 90th and 99th percentiles and maximum of the per-grain volume error, as a fraction of the
mean initial grain volume, over the common window $t\le50$ (150 and 1,000 grains: three runs pooled;
5,000 grains: one run); dashed: the $0.05$ criterion on the maximum. (b)~Maximum error of each 150-
and 1,000-grain run to $t=135$; filled markers: first time above $0.05$. (c)~Extinction-time offsets
of matched grains, PINN-Phase minus reference, against the reference extinction time; shading:
tolerance $\max(100,\,0.1\,s_{\mathrm{ref}})$ steps; open triangles: events straddling the horizon.
Right: time to $N/N_0=0.60$, PINN-Phase minus reference.}{fig:fidelity}

\subsection{What matched-state statistics detect}
\label{sec:results:controls}

Mean size against time, face statistics and normalized size distributions are the usual measures
for validating large-scale grain-growth models~\cite{tian2026primme}. The time-rescaling controls of
Section~\ref{sec:methods:population} show which question each of them answers
(Figure~\ref{fig:controls}). Compared at the same state, the controls cannot be told apart from the
reference. The mean number of faces against $\bar R^2$ coincides for all four trajectories. At
$N/N_0=0.60$, the generalized Hillert exponents are $4.253$ (reference), $4.251$ (PINN-Phase), and
$4.248$ and $4.247$ (controls), with Kolmogorov--Smirnov distances to the reference of at most
$0.0025$. Against time, the controls run ahead: $N/N_0=0.60$ is reached at $t=47.3$ by the reference,
$47.4$ by PINN-Phase, and $45.9$ and $43.0$ by the controls. The per-grain comparison separates them
most clearly. At $t=30$, the 90th percentile of the per-grain volume error is $0.0039$ for PINN-Phase,
against $0.033$ and $0.109$ for the controls. Over 1,921 extinctions matched within $t\le45.45$, the
median extinction offsets are $+0.03$, $-0.80$ and $-2.50$ time units. Over the full window, the
growth rates of PINN-Phase and reference grains correlate with a Pearson coefficient of $0.99995$
(93,529 paired grain-time records), and $98.1\%$ of 92,951 paired face counts are identical.

A model with a 3\% clock error therefore reproduces the expected sequence of grain-growth states,
with the correct topology and size distribution, while placing every state at the wrong time.
State-matched statistics establish that the path is right. Time-resolved and grain-resolved
comparisons are needed to establish that it is traversed at the right rate and that individual grains
follow their reference histories.

\skelfig{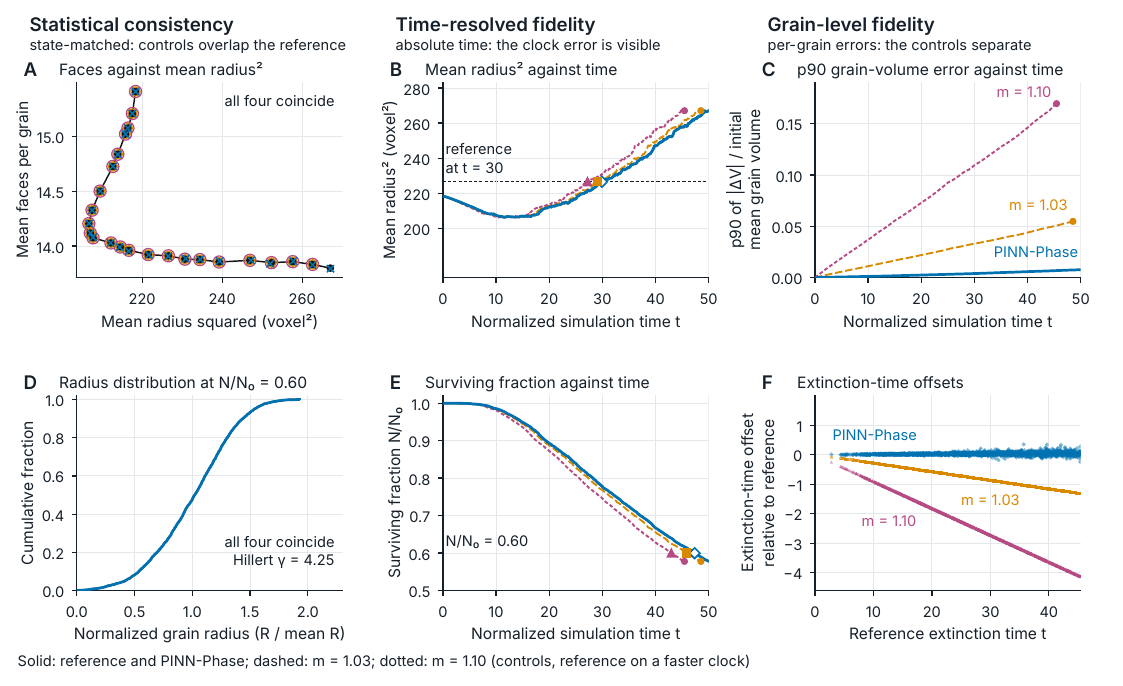}{%
\footnotesize\textbf{What different validation measures detect (one 5,000-grain microstructure).}
The controls replay the reference on a 3\% ($m=1.03$) or 10\% ($m=1.10$) faster clock, up to
$t=48.54$ and $45.45$; they are time rescalings, not trained models. (a,~d)~Compared at the same
coarsening state, face statistics and the normalized radius distribution coincide for all four by
construction. (b,~e)~Against time, mean grain size and surviving fraction reveal the clock error.
(c,~f)~Per-grain volume errors and extinction-time offsets separate PINN-Phase from both controls.
Solid: reference and PINN-Phase; dashed: $m=1.03$; dotted: $m=1.10$.}{fig:controls}

\subsection{Consequence for a virtual isothermal hold}
\label{sec:results:screening}

The difference between a correct path and a correct rate matters when a simulation is used to make
a processing choice. In the virtual isothermal hold (Section~\ref{sec:methods:population}), each
model chooses the shortest hold at which the mean grain volume has grown by a target factor
(Figure~\ref{fig:screening}). In the prospective 5,000-grain case, PINN-Phase reaches the four targets ($Q\ge1.01$, $1.02$, $1.25$
and $1.50$) at most $0.22\%$ later than the reference. It never stops before the reference has reached
the target. For $Q\ge1.50$, the reference stops at $t=40.1$, PINN-Phase at $40.2$, and the $m=1.03$
and $m=1.10$ clocks at $39.0$ and $36.5$.

Across all 34 case-target decisions, the PINN-Phase time-to-target error stays within $0.29\%$. One
stop is premature: a retrospective 1,000-grain decision for $Q\ge1.50$, one 0.1-unit grid step early
($t=38.6$ instead of $38.7$). Both controls stop prematurely in all 34 decisions. The $m=1.10$ clock
misses the time by up to $9.1\%$, outside the 5\% tolerance. The $m=1.03$ clock stays within it
(largest error $2.9\%$), yet it still stops before the target is reached in every decision. A
tolerance on timing error and the attainment of the target are therefore different requirements, and
a small systematic clock error fails the second consistently. Because the target follows the mean
volume, the test probes global coarsening time, which a coarsening curve against absolute time would
also reveal. It shows the decision-level value of correct kinetic timing and complements, rather than
replaces, the grain-resolved validation above (per-decision values in Supplementary
Section~\ref{sm:screening}).

\skelfig{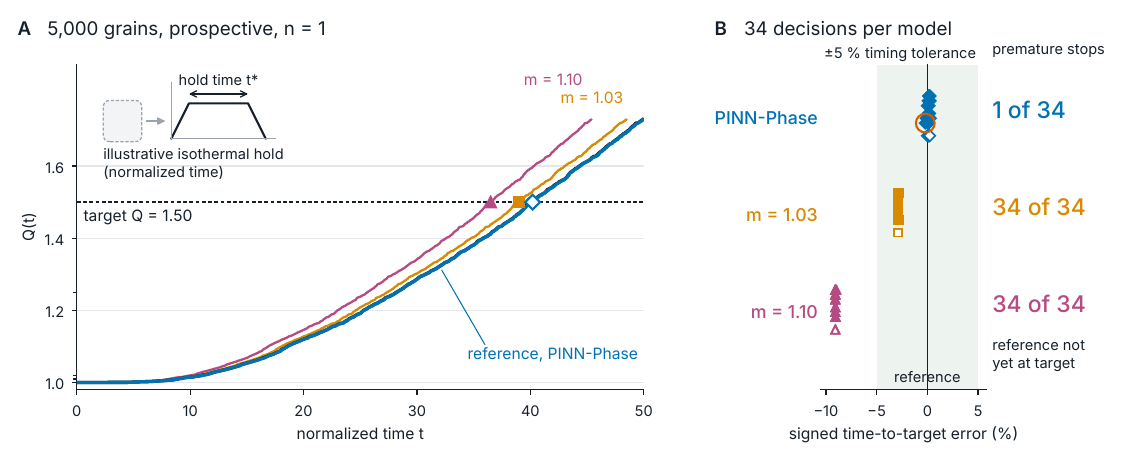}{%
\footnotesize\textbf{Virtual isothermal hold in normalized simulation time.} Each model chooses
the first time, on a 0.1-unit grid, at which the mean alive grain volume reaches a target multiple
$Q$ of its initial value; the inset is illustrative, and no thermal field is simulated. Controls as
in Figure~\ref{fig:controls}. (a)~Prospective 5,000-grain case, $Q=1.50$: stop times $40.1$
(reference), $40.2$ (PINN-Phase), $39.0$ and $36.5$ (controls). (b)~Signed time-to-target error for
34 correlated case-target decisions per model (open: the four prospective ones), with the $\pm5\%$
timing tolerance shaded; right, premature stops, at which the reference has not yet reached the
target (ring: the single premature PINN-Phase stop).}{fig:screening}

\FloatBarrier
\end{revision}

\begin{revision}
\section{Discussion}
\label{sec:discussion}

\subsection{Grain-resolved validation for numerical microstructure exploration}
\label{sec:discussion:validation}

A model can reproduce the expected stages of grain growth while predicting incorrectly when they
occur. The time-rescaling controls make this concrete. A trajectory on a clock 3\% too fast has the
reference topology, face statistics and normalized size distribution at every matched coarsening
state. Yet its grains have the wrong sizes at any given time, its extinctions come early, and it ends
the virtual hold before the target is reached in every decision (Sections~\ref{sec:results:controls}
and~\ref{sec:results:screening}). Statistics compared at matched states therefore show that a model
follows the right microstructural path. Whether it follows that path at the right rate, and whether
each grain follows its own history, requires comparisons in time and grain by grain. Mean size
against absolute time detects a clock error to the resolution of the kinetic comparison; per-grain
volumes and extinction times detect it directly.

Large learned grain-growth models have shown that deployment at scale is possible, with strong
statistical agreement against their reference models~\cite{tian2026primme}. The present evidence
addresses a complementary requirement. With unchanged weights, the phase-local realization reproduces
which grains disappear and when, from 150 to 5,000 grains. Its typical per-grain error does not grow
with the number of grains through the common window, and the hold time it selects agrees with the
reference. The comparison is made against one deterministic projected-obstacle reference. It does not
rank methods built on different reference models, and trajectory divergence under a stochastic
reference is partly intrinsic to that reference. For numerical materials exploration, this motivates
a layered validation approach: global kinetics and grain-growth statistics establish
population-level behavior, while grain-volume histories and extinction events test fidelity at the
level of individual grains. The screening decision must then be tested on its own terms.

\subsection{A structurally admissible neural time integrator for multiphase fields}
\label{sec:discussion:established}

\drev{PINN-Phase combines three functions that must remain compatible during long autoregressive
evolution. A physics-informed rate model is evaluated on the evolving predicted field, a neural time
integrator retains phase identity through topology change, and the state update enforces local
phase-fraction constraints after every step. Across the benchmark ladder, this combination remains
admissible as grains disappear. The references are direct integrations of the same discretized
operator, so the errors measure numerical fidelity, not material validation.} \drev{In the hybrid
realization, projection of the Allen--Cahn rate makes the residual tangent to the unit-sum manifold,
phase-mean removal makes the neural increment phase-sum neutral, and clip-and-renormalize constrains
the carried state. The site-local and convolutional-recurrent branches together are substantially
more accurate than either alone, although the larger parameter count of the hybrid prevents
attributing this to branch composition alone.}

\drev{Across the multigrain benchmarks, the hybrid formulation preserves admissible phase fields during
grain disappearance, but agreement depends on both the case and the length of the rollout. Long
trajectories can recover the correct surviving grains while retaining errors in boundary placement or
extinction time. The prospective native-grid tests likewise show that strong field agreement does not
guarantee exact event timing in every case (Figures~\ref{fig:voronoi25} and~\ref{fig:n128transfer};
Supplementary Sections~\ref{sm:c} and~\ref{sm:l}). These outcomes justify reporting field,
grain-identity and event-time measures separately.}

\subsection{Relation to physics-informed solvers and microstructure surrogates}
\label{sec:discussion:literature}

\drev{Conventional PINNs represent solutions as coordinate networks constrained by
differential-equation residuals~\cite{raissi2019,karniadakis2021}. Temporal decomposition and time
marching extend them to longer transients~\cite{penwarden2023causal,meng2020ppinn,jagtap2020xpinn}.
Convolutional-recurrent physics-informed models advance an entire field recurrently, with boundary
conditions encoded structurally~\cite{REN2022114399}. PINN-Phase adopts this viewpoint for an
$N$-component phase-fraction field whose admissibility and active phase set change during evolution.
PINNs-MPF uses phase-wise coordinate networks with soft phase-sum
enforcement~\cite{elfetni2025pinnsmpf}. The hybrid realization instead carries all channels in one
shared autoregressive field model and imposes admissibility in the state update, and the phase-local
realization shares its learned parameters across phases. These are complementary strategies, not
ranked implementations.}

\drev{Other learned microstructure models use precomputed simulation trajectories for
training~\cite{hu2022accelerating,montes2021accelerating,oommen2022learning,qin2024graingnn,zhou2026voxelflow,tian2026primme}.
These include methods that extend to very large three-dimensional domains or combine learned
forecasts with physical correction. PINN-Phase follows a different training route and retains
continuous constrained phase fields. It learns from the governing residual on its own evolving state,
without later phase-field states in the loss or checkpoint selection, and it retains phase identity
through topology change. In the $N=25$ and $N=64$ formulation, it is also} \crev{exactly equivariant to
phase permutations and integer periodic translations of the grid.} \drev{Since the training data,
governing operators and state representations differ, these studies do not form a controlled accuracy
ranking. The optional graph conditioner is a topology-sensitive extension with modest, case-dependent
effects (Supplementary Figure~\ref{fig:gnn}), not a universal improvement.}

\subsection{\drev{Scope and implications for microstructure exploration}}
\label{sec:discussion:scope}

PINN-Phase offers complementary forms of reuse within the tested numerical models. The hybrid
realization generalizes to unseen initial conditions and evolves beyond its represented training
horizon. The independently trained phase-local realization reuses the same learned weights as domain
size and grain population increase. These results establish complementary capabilities, not transfer
between the two architectures or their different physical operators. Because the matched comparator
trained on reference-trajectory states also meets its criteria, the results show that training on
the model's own states is viable, not that it causes the transfer.

The grain-resolved tests also define an accuracy envelope. Typical grain-volume errors remain small
across the tested populations over the shared comparison window, while a few grains develop larger
deviations at later times. Event timing and extreme grain-volume errors therefore remain essential
checks, even when average coarsening statistics agree. The single 5,000-grain microstructure provides
a paired large-scale demonstration over its reported horizon; it does not establish population-level
reliability across many such initial conditions.

The virtual isothermal-hold study illustrates why this distinction matters for numerical screening. A
model that advances through the expected grain-growth states at the wrong rate can recommend a
premature stopping time. The present target depends on the mean volume of surviving grains, so correct
global kinetics, rather than grain identity alone, are sufficient to evaluate that particular decision.
Grain-resolved tests provide a complementary safeguard when the evolution of specific grains is
important. Within these limits, the unchanged phase-local rule can support repeated numerical
screening of related microstructures, with direct integration retained to verify selected cases.
Because the update is local and grain remapping keeps the number of channels in use nearly constant,
the memory of a run follows its grid size rather than its number of grains.

These findings remain limited to normalized-time simulations with specified isotropic phase-field
operators. The phase-local transfer tests use fixed grid spacing and physical coefficients; they do
not establish transfer across materials, resolutions or boundary-property laws. The learned parameter
count is independent of grain number, although the full numerical update remains coupled across
phases. Experimental calibration, for example by data assimilation~\cite{yamanaka2019enkf},
anisotropic evolution~\cite{moelans2022anisotropic}, longer large-population horizons and the
computational efficiency of the learned step require separate studies. Detailed qualification records
remain in the Supplementary Material.

\end{revision}

\section{Conclusion}
\label{sec:conclusion}

PINN-Phase learns autonomous updates for continuous multiphase fields from a physical operator
evaluated on the model's own predicted states. A constrained update preserves admissible phase
fractions as interfaces evolve and grains disappear. The hybrid realization generalizes prospectively
to unseen two- and three-dimensional microstructures; it meets all predefined criteria in seven of
eight unseen two-dimensional cases and both stress cases. The separately trained phase-local
realization reuses its 2,160 learned parameters unchanged from eight grains at $48^3$ to 5,000 grains
at $416^3$. It reproduces which grains disappear and when, although a few large grains exceed the
strict volume criterion at late times.

The grain-resolved and virtual-screening tests show why kinetics and individual-grain histories
should be assessed alongside aggregate statistics. A 3\% clock error passes statistics compared at
matched coarsening states, yet it ends a virtual hold early in every decision, whereas PINN-Phase
stays within $0.3\%$ of the reference time. These findings apply to the specified numerical operators
in normalized time. Material-specific calibration, anisotropic growth and longer horizons at the
largest populations remain to be established.
\par
{\footnotesize\begin{revision}
\section*{Declarations}

\paragraph{Declaration of competing interest}
The authors declare that they have no known competing financial interests or personal relationships
that could have appeared to influence the work reported in this paper.

\paragraph{Funding}
This work was supported by the German Federal Ministry of Research, Technology and Space through
the project ``StStG CTC -- Wissen schafft Perspektiven f\"ur die Region: Center for the
Transformation of Chemistry gGmbH'' (funding reference 03WSP1573).

\paragraph{Data and code availability}
The source code, released model weights, benchmark packages and evaluation records, including those
of the 150-, 1,000- and 5,000-grain evaluations, the time-rescaling controls and the virtual
isothermal hold, are available under the BSD 3-Clause License at
\url{https://github.com/SFETNI/PINN-Phase} (release v4).

\paragraph{Acknowledgments}
The first author acknowledges Prof. Reza Darvishi Kamachali for early scientific discussions on
combining machine-learning approaches with multiphase-field simulation, which contributed to the
broader research context in which PINN-Phase was developed.

\end{revision}
\par}
\FloatBarrier

\bibliographystyle{elsarticle-num}
\bibliography{references}

\end{document}